\documentclass[twocolumn,trackchanges]{aastex7}

\usepackage{xspace}
\usepackage{mhchem}
\usepackage{mathrsfs} 
\usepackage{hyperref}

\newcommand \kzz{$K_{zz}$}
\newcommand \fsed{$f_{\mathrm sed}$}
\newcommand {\um}{$\mathrm{\,\mu m}$}
\newcommand \rjup{R$_{\mathrm Jup}$}
\newcommand \mjup{M$_{\mathrm Jup}$}
\newcommand{\target}{WD0806$\,$b\xspace}
\newcommand{\red}[1]{\textcolor{red}{#1}}
\newcommand{\teff}{T$_{\mathrm{eff}}$}
\newcommand{\logkzz}{$\log (K_{zz})$}
\newcommand{\rev}[1]{\textcolor{black}{#1}}
\begin{document}

\title{JWST spectral retrieval of cold directly imaged planet WD0806b and the first measurement of altitude-dependent \kzz\ in exoplanet atmospheres}

\author[0000-0003-1487-6452]{Ben W.P. Lew}
\affiliation{Bay Area Environmental Research Institute, Moffett Field, CA 94035, USA}
\affiliation{NASA Ames Research Center, Moffett Field, CA 94035, USA}
\email{weipeng.lew@nasa.gov,lew@baeri.org}

\author[0000-0002-6730-5410]{Thomas Roellig}
\affiliation{NASA Ames Research Center, Moffett Field, CA 94035, USA}
\email{thomas.l.roellig@nasa.gov}

\author[0000-0003-1240-6844]{Natasha E. Batalha}
\affiliation{NASA Ames Research Center, Moffett Field, CA 94035, USA}
\email{natasha.e.batalha@nasa.gov}

\author[0000-0002-0413-3308]{Nicholas F. Wogan}
\affiliation{NASA Ames Research Center, Moffett Field, CA 94035, USA}
\email{nicholas.f.wogan@nasa.gov}

\author[0000-0002-8963-8056]{Thomas Greene}
\affiliation{NASA Exoplanet Science Institute, Infrared Processing and Analysis Center (IPAC)}
\email{tgreene@ipac.caltech.edu}

\author[0000-0002-5251-2943]{Mark S. Marley}
\affiliation{Lunar and Planetary Laboratory, University of Arizona, Tucson, AZ 85721, USA}
\email{marksmarley@arizona.edu}

\author[0000-0002-9843-4354]{Jonathan J. Fortney}
\affiliation{University of California Santa Cruz, 1156 High Street, Santa Cruz, CA 95064, USA}
\email{jfortney@ucsc.edu}

\author[0000-0002-0834-6140]{Jarron Leisenring}
\affiliation{Steward Observatory, University of Arizona, Tucson, AZ 85721, USA}
\email{jarronl@arizona.edu}

\author[0000-0002-6773-459X]{Doug Johnstone}
\affiliation{NRC Herzberg Astronomy and Astrophysics, 5071 West Saanich Rd, Victoria, BC, V9E 2E7, Canada}
\affiliation{Department of Physics and Astronomy, University of Victoria, Victoria, BC, V8P 5C2, Canada}
\email{Doug.Johnstone@nrc-cnrc.gc.ca}

\author[0000-0003-1863-4960]{Matthew De Furio}
\affiliation{Department of Astronomy, University of Texas at Austin,2515 Speedway, Stop C1400 Austin, Texas 78712-1205, USA}
\email{matthew.defurio@austin.utexas.edu}

\author[0000-0003-0786-2140]{Klaus Hodapp}
\affiliation{University of Hawaii, Hilo, HI,96720, USA}
\email{hodapp@hawaii.edu}

\author[0000-0002-5627-5471]{Charles Beichman}
\affiliation{NASA Exoplanet Science Institute, Infrared Processing and Analysis Center (IPAC)}
\affiliation{Jet Propulsion Laboratory, California Institute of Technology, Pasadena, CA 91125, USA}
\email{chas@ipac.caltech.edu}


\author[0000-0002-7893-6170]{Marcia Rieke}
\affiliation{Steward Observatory, University of Arizona, Tucson, AZ 85721, USA}
\email{mrieke@as.arizona.edu}


\begin{abstract}
\target{} is a rare exoplanet companion orbiting a white dwarf, currently with a projected orbital distance of 2500\,au. The Spitzer mid-IR photometry suggests that the temperature is as cold as 350K, making it one of the coldest directly imaged exoplanets. In this paper, we present the Near-infrared Camera (NIRCam)  F150W2, F200W, F356W, and F444W broadband photometry and a 3--5\um\ Near-Infrared spectroscopy (NIRSpec) G395M spectrum obtained with the James Webb Space Telescope (JWST).  
We develop a new retrieval framework based on the open-source PICASO software that includes additive and multiplicative systematic parameters.
Our retrieval results reveal bounded abundances of \ce{H2S}, \ce{CO2}, \ce{CO}, \ce{NH3}, \ce{H2O}, and \ce{CH4}. 
We present a new chemical analysis framework that utilizes retrieved abundances to measure altitude-dependent eddy diffusion coefficients (\kzz) at multiple quenched pressures. We find that the eddy diffusion coefficients decrease from around $10^4$ to $10^2$ $\rm cm^2/s$ as the atmospheric pressure decreases from from 50 to 20 bars.
To our knowledge, this is the first study to report altitude-dependent vertical mixing (or, equivalently, quenched-species-dependent vertical mixing) based on the measured molecular abundances of \ce{CO}, \ce{CH4}, and \ce{CO2}.
With the 1--21\um NIRCam, NIRSpec and the previously published MIRI data, we measure the bolometric luminosity to be log(L/L$_{\odot}$) = $-6.75\pm0.01$ and derive the mass to be $8\pm 1  \mathrm{M_J}$.
The retrieval results suggest that \target has an elevated C/O ratio of 0.76, or  1.3$\times$ solar, sub-solar metallicity ([M/H ]= -0.25), and a nearly solar C/S ratio (1.17x solar).

\end{abstract}



\section{Introduction} \label{sec:intro}
Directly imaged planet systems provide key insights into how giant planets form and evolve.
The observed bolometric luminosity and the emission spectral features inform the mass, radius, and atmospheric composition of these objects.
The age of the host star provides prior knowledge about the age of the planet, allowing us to break the degeneracy between fundamental parameters such as mass and age in the atmospheric characterization.
By comparing the host star and exoplanet abundances, we can test if the star and planets share similar or different elemental compositions, and investigate the potential planet formation and evolution pathways.

To date, there are about 90 directly imaged exoplanets and planetary-mass companions \footnote{\url{https://exoplanetarchive.ipac.caltech.edu}}, mostly discovered with ground-based telescopes. 
With the unprecedented mid-infrared sensitivity of James Webb Space Telescope (JWST), we can expand this population to previously inaccessible cooler targets.
The recent discovery of Eps Ind Ab \citep{matthews2024}, and the detection of WD 1856 b \citep{limbach2025}  showcase  JWST's capability in discovering and characterizing emission from cold giant planets.
Meanwhile, JWST observational studies of Y-dwarfs, which are excellent exoplanet analogs that share temperature and mass with cold directly imaged planets, unveil exciting yet sometimes puzzling atmospheric chemistry and physical properties in these cold ($< 500\,$K) atmospheres. 
For example: \ce{PH3} is absent in almost all observations of brown dwarfs and exoplanets \citep[e.g.,][]{beiler2023,burgasser2024,rowland2024}.
The detection of \ce{H2S} \citep[e.g][]{lew2024} enables us to study sulfur chemistry in giant planet atmospheres.
The \ce{CH4} emission features that hint of aurora-like or other heating mechanisms in their upper atmospheres \citep[][]{faherty2024}. Finally, the \ce{CO2} abundances are higher than expected compared to atmospheric models \citep[e.g.,][]{beiler2024,mukherjee2024}.
These findings highlight that it is critical to study the atmospheric chemistry and physical processes to accurately interpret the spectral signature of cold giant planets.

Cold directly imaged planets are rare, and there are even fewer giant planets found to be orbiting a white dwarf. 
The exoplanets orbiting these late stage of stellar evolution stars prompt intriguing questions: how does exoplanet survive and evolve with the host star?; did the ejected stellar winds of the progenitor host star leave detectable imprints in the observed planetary atmosphere?
To answer these questions, the WD0806 system was targeted in the JWST GTO Program PID 1276 to characterize the  atmosphere of its planetary companion. WD0806b is a unique benchmark system because it orbits a white dwarf and it is the second coldest directly imaged exoplanet behind Eps Ind Ab \citep{matthews2024}. As one of the coldest companions, \target{} is also invaluable to be compared with cold Y-Y dwarf binary systems \citep[e.g.,][]{calissendorff2023,defurio2025} in studying the potential link between formation pathways and atmospheric composition.

In the following sections, we first introduce the critical role of non-equilibrium chemistry in giant planet atmospheres in Section \ref{sec:kzzintro}. We then summarize the properties of the WD0806 system in Section \ref{sec:targetintro} and literature studies in Section \ref{sec:voyer2025}. In Section \ref{sec:reduction}, we describe the observing setup, data reduction, and the estimated systematic errors in our JWST observations. In the results section, we first present the measurement of bolometric luminosity and mass estimate in Section \ref{sec:lbol}, then describe the forward modeling and atmospheric retrieval results in Section \ref{sec:spectral_modeling}, and finally present the new framework for deriving species dependent eddy diffusion coefficients (\kzz) in Section \ref{sec:kzz}.
We discuss the physical interpretation of our key results, caveats of the modeling tools, and the implications for cold giant planet atmospheres in Section \ref{sec:discussion}.

\subsection{Non-equilibrium chemical abundances as a probe of dynamical mixing variation in an atmosphere}\label{sec:kzzintro}

Strong absorption from molecules like \ce{NH3} and \ce{CH4} shape the main spectral features of cold giant planets.
These molecules, along with other minor species such as \ce{CO2} and CO, are sensitive to non-equilibrium chemistry.
The molecular abundance is in chemical non-equilibrium when the limiting chemical reaction timescale $\tau_{chem}$, exceeds the dynamical mixing timescale $\tau_{mix}$, i.e.,  $\tau_{chem}>\tau_{mix}$.
Since atmospheres generally cool with increasing altitude, and reaction rates are generally directly proportional to T and P, $\tau_{chem}$ increases as an atmospheric parcel is mixed upward.
In the non-equilibrium regime where $\tau_{chem}>\tau_{mix}$, the dynamical mixing homogenizes the molecular abundance across altitudes, driving the molecular abundances to be out of an equilibrium state.
The highest pressure (lowest altitude) at which the disequilibrium chemistry occurs is often known as the quenched pressure, $P_{quenched}$. 
At $P_{quenched}$, were the chemical timescale is equal to the dynamical mixing timescale.
At pressures below $P_{quenched}$, the molecular abundance is homogenized.
The dynamical timescale is often parameterized by $L^2/$\kzz, where L is the mixing length scale and \kzz is the eddy diffusion coefficient.
The observed abundances of a molecule that deviates from the chemical equilibrium model can thus be used to estimate the dynamical mixing strength at the quenched pressure.

There are multiple atmospheric models that include \kzz parameters in modeling non-equilibrium chemical processes \citep[e.g.,][]{karalidi2021,mukherjee2024,phillips2020}. In these models, \kzz is included as an altitude-independent parameter to explore the impact of non-equilibrium chemistry on the emission spectra.
Yet, dynamical mixing or \kzz is known to vary with altitudes and affect the temperature-pressure (TP) profile, cloud formation, and chemical compositions \citep[e.g.,][]{ackerman2001,hubeny2007,freytag2010, showman2013,mukherjee2022, tan2022}.
Observational studies of the \kzz variation within an atmosphere is key to test the current non-equilibrium chemistry modeling framework and to better quantify the impact of dynamical mixing on the atmospheric thermal and chemical composition.

\subsection{WD0806 and \target{} properties}\label{sec:targetintro}

\target was discovered by \citet{luhman2011} based on the Spitzer Infrared Array Camera (IRAC) [4.5\,\um] images taken at two epochs. 
The host, WD 0806-661 (L97-3), is a DQ white dwarf with a mass of 0.62 $\pm 0.03$ M$_{\odot}$ \citep{subasavage2009}. 
Based on the white-dwarf initial-final mass relation \citep{catalan2008}, \citet{rodriguez2011} we estimate the progenitor mass to be 2.0$\pm 0.3$ $M_{\odot}$ and the age to be 1.5--2.7 Gyr. 
\cite{luhman2012} reported Spitzer IRAC [3.6\um] photometry and an upper limit of J-band photometry. 
They used the white dwarf initial-final mass relations \citep{kalirai2008,williams2009} and the isochrones tables by \citet{girardi2002} to estimate the age of WD0806 to be 2 $\pm 0.5$ Gyr.
 Follow-up near-infrared observation by \citet{luhman2014} used the Wide Field Camera 3 (WFC3) onboard of the Hubble Space Telescope (HST) measured an F110W magnitude of 25.70$\pm$0.08 for \target{}.

\citet{voyer2025} published the JWST MIRI Low-Resolution Spectroscopy (LRS) spectra and F1200, F1500, F1850, and F2100 band photometry of \target. 
In \citet{voyer2025}, they used the TAUREX3 atmospheric retrieval code \citep{refaie2021,refaie2022} and explored various retrieval setups to fit the MIRI-LRS spectra.
Based on the retrieval results, they reported the detected abundances of \ce{NH3}, \ce{H2O}, \ce{CH4} and placed upper limits on \ce{CO2} and CO abundances, respectively.
They also explored various setups of retrieval, including multi-layer \ce{NH3} and \ce{H2O} abundances and cloud opacity. 
Among the explored retrieval setups (see Section 2.4 in \citealt{voyer2025}), they found no evidence of cloud opacity nor a two-layer \ce{H2O} abundance. 
However, they found that the retrieval with two-layer \ce{NH3} is preferred over uniform-\ce{NH3} retrieval with a Bayesian evidence difference of 23.5.
Their retrieval mass of 0.45--1.75 $M_J$ is significantly lower than evolutionary model predictions and implies \target to be a very young object (60 - 180 Myr old) that is much younger than the white dwarf age (2$\pm$0.5 Gyr), which would indicate that either  \target{} formed after the progenitor star or was gravitationally captured \citep[e.g.,][]{varvoglis2012,goulinski2018,li2019,yu2024}. 
We further discuss the results of \citet{voyer2025} in Section \ref{sec:lbol}.

\section{Observations and Data Reduction}\label{sec:reduction}
The JWST NIRSpec and NIRCam observation of WD 0806 b was part of the JWST GTO Program PID 1276 (P.I Pierre Olivier Lagrange).
The NIRSpec Integral Field Unit (IFU) spectroscopic observations were executed using the NRSIRS2RAPID readout mode with the G395M grating combined with a F290LP filter, whose wavelength range spans from 2.87 to 5.10 \um. 
The IFU observation consisted of a 4-POINT-NOD dithering pattern over four exposures.
Within each exposure, there were 4 integrations and 26 groups in each integration. 
The total exposure time is 6302 seconds.

\begin{figure*}
\centering\includegraphics[width=.8\textwidth]{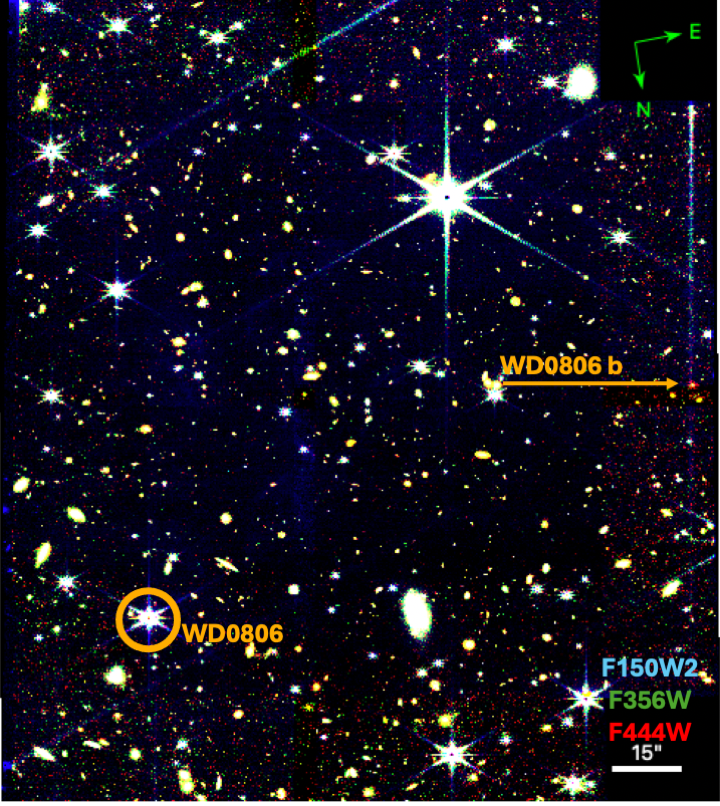}
\caption{The RGB-stacked colored image with the three NIRCam filters: F150W2 (blue), F356W (green) and F444W (red) using DS9 software \citep{ds9}.The location of white dwarf WD0806 is shown in the orange circle and the location of WD0806 b is illustrated by the orange arrow.}\label{fig:nircam}
\end{figure*}

We also obtained four broadband NIRCam photometry images with the F150W2, F200W, F356W, and F444W filters.
Each of the NIRCam broadband images had an exposure time of 934 seconds using the SHALLOW4 readout mode.
Each visit contained three dithers. There were five groups within each integration and one integration in each dither observation.
We present the RGB-stacked NIRCam images in Figure \ref{fig:nircam}. Based on the NIRCam images and source catalog from the JWST pipeline, we did not find any objects in the field of view that shared similar or redder colors than \target{}.

For the NIRSpec data reduction, we downloaded the stage-0 data (uncal.fits, \rev{SDP\_VER = "2023\_2a"}) and reran the STScI pipeline \rev{(CAL\_VER = 1.14.1.dev62+g6c29b41d)}  with the CRDS context 1258.pmap.
After the stage-2 pipeline, we used a customized module that removed outliers that exhibited a strong spatial gradient and flux with the neighboring pixels through a sigma-clipping method. 
At the stage-3 of the pipeline, we used a 3x3 kernel in the outlier-detection step when combining the 4-dithering IFU data into the IFU datacube. The spectral extraction was done on the stage-3 pipeline product with an aperture radius of 4.5 pixels, the same aperture used by NIRSpec team for absolute flux calibration.

\subsection{Aperture photometry for the NIRCam images}
For the NIRCam images we used the standard JWST pipeline product processed under CRDS context pmap 1256. In order to maximize the signal-to-noise we adjusted the aperture radius such that the aperture radius was larger for a brighter source. 
The aperture radii used were 4, 3, 12, and 20 pixels for the F150W2, F200W, F356W, and F444W images respectively.
Upon visual inspection of the background structure, we adopted different strategies for the background subtraction:
For the 150W2 image, we used a nearby row-averaged background to remove the row-by-row background structure; 
For the F356W and F444W images, we used the median flux level of a source-free box region;
For the F200W image, we used a column-and-row averaged master background to remove the row- and column-dependent background in the image.
Finally, we performed the absolute flux calibration using the encircled energy function for the NIRCam to calibrate the flux extracted using the different aperture radii.
 The \target absolute-flux calibrated NIRCam aperture photometry measurements in units of ergcm$^{-3}$s$^{-1}$ are $1.10 \pm 0.05 \times 10^{-12}$ for F150W2, $3.41 \pm 0.51 \times 10^{-13}$ for F200W, $ 6.48 \pm 0.07 \times 10^{-12}$ for F356W, and $4.43 \pm 0.01 \times 10^{-11}$ for F444W.
We cross-validated the NIRCam F356W and F444W photometry with the NIRSpec/G395M spectrum.
Based on the integrated flux from the G395M spectrum, we find that the integrated fluxes are consistent with the NIRCam F356W and F444W photometry within 1 $\sigma_{full}$, which is the total uncertainty that includes, pipeline uncertainty, fitted noise floor, and fitted flat-field uncertainties as described in Section \ref{sec:noise}.

\subsection{\rev{Measurements of the Full Width Half Maximum of the instrument line profile}}
Knowing the instrumental spectral resolution is important to properly convolve the model spectra to fit to the observational data. The spectral resolution of the NIRSpec grating is wavelength-dependent. 
To our knowledge, there is no published in-flight NIRSpec spectral resolution measurement.
Similar to \citet{lew2024}, we used a Gaussian model to measure the Full Width Half Maximum (FWHM) of an unresolved emission line at 3.4\,\um\ of a planetary nebula obtained under the NIRSpec calibration program (PID 1128).  
We approximated the wavelength-dependence in the published NIRSpec/G395M dispersion curve \footnote{\url{https://jwst-docs.stsci.edu/jwst-near-infrared-spectrograph/nirspec-instrumentation/nirspec-dispersers-and-filters}} with a linear regression. 
We then scaled the linear dispersion curve to match the measured FWHM at 3.4\um\  as the \rev{approximate} instrumental broadening profile to convolve the model spectra during the spectral fitting of grid models and retrievals in Section \ref{sec:spectral_modeling}.

\begin{figure*}
    \centering
    \includegraphics[width=1\linewidth]{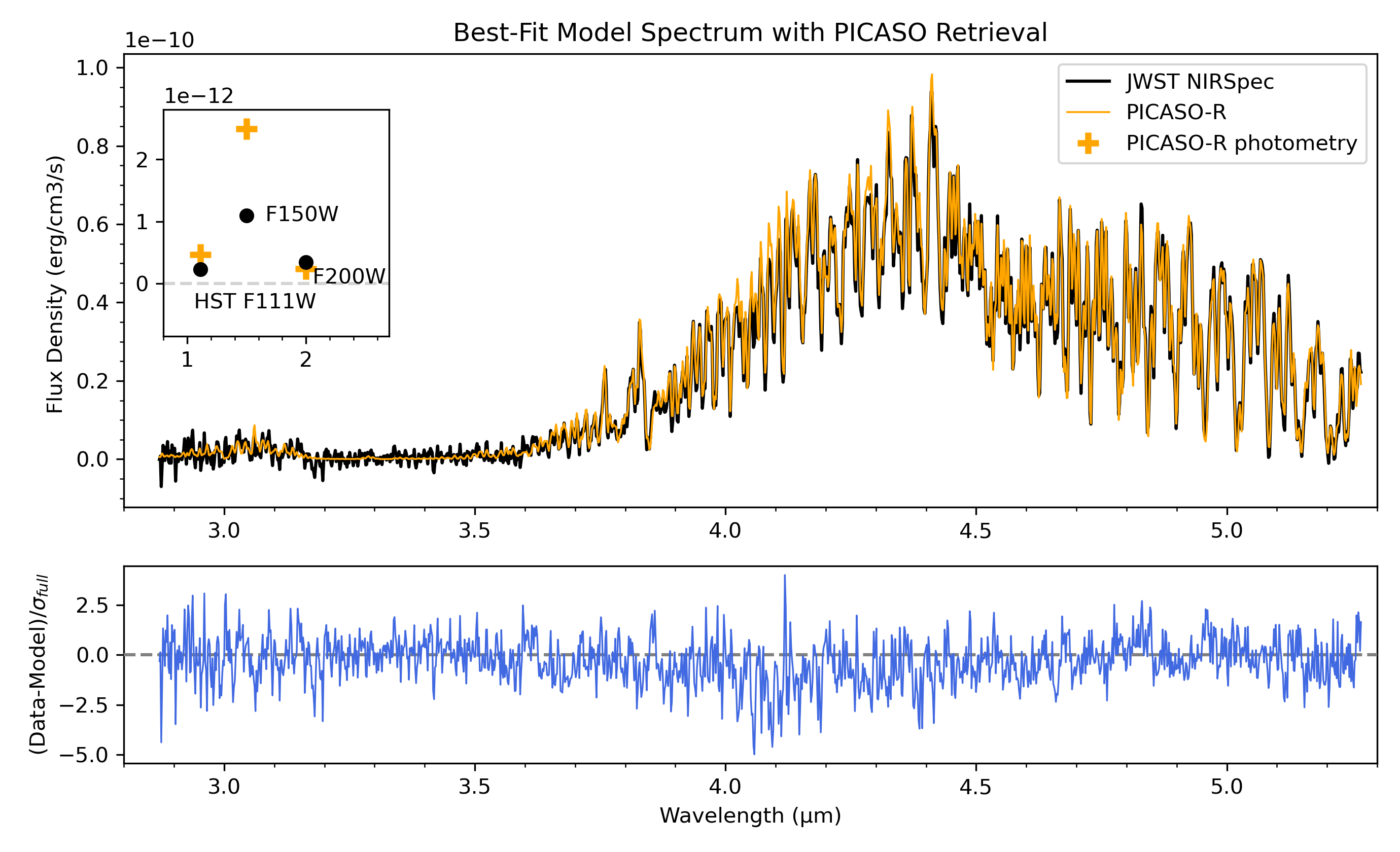}
    \caption{\textbf{Top panel}: the solid black and orange lines show the NIRSpec/G395M spectrum and the best-fit retrieval spectrum respectively. \rev{The inset plot shows the JWST NIRCam F150W2, F200W, and the HST F110W broadband photometry, which are plotted in black dots}. The integrated band photometry of the best-fit retrieval spectrum area plotted in orange plus signs. \textbf{Bottom panel}: the difference between the NIRSpec spectrum and the best-fit retrieval that is normalized by the full uncertainty ($\sigma_{full}$), which includes NIRSpec/G395M data uncertainty and the two the marginalized uncertainty parameters described in Section \ref{sec:noise}.}\label{fig:spec}
\end{figure*}

The reduced NIRSpec spectrum is shown in Figure \ref{fig:spec}, together with the retrieval results we obtained with PICASO (Section \ref{sec:picasor}).

\section{Results}
In this results section, we first present our derivation of the bolometric luminosity and mass using brown dwarf evolutionary models in Section \ref{sec:lbol}. The inferred bulk properties such as mass and radius from the evolutionary model provide physical constraints in the following spectral modeling presented in Section \ref{sec:spectral_modeling}. With the characterized molecular abundances, we present the inferred elemental abundance ratios and derive the dynamical mixing strength based on a non-equilibrium chemistry model in Sections \ref{sec:elementratio} and \ref{sec:kzz}. 
These analyses showcase the comprehensive physical and chemical picture for one of the coolest exoplanet atmospheres observed with a state-of-the art JWST dataset.

\subsection{Bolometric luminosity and bulk properties}\label{sec:lbol}
\begin{figure*}
    \centering
    \includegraphics[width=1.0\linewidth]{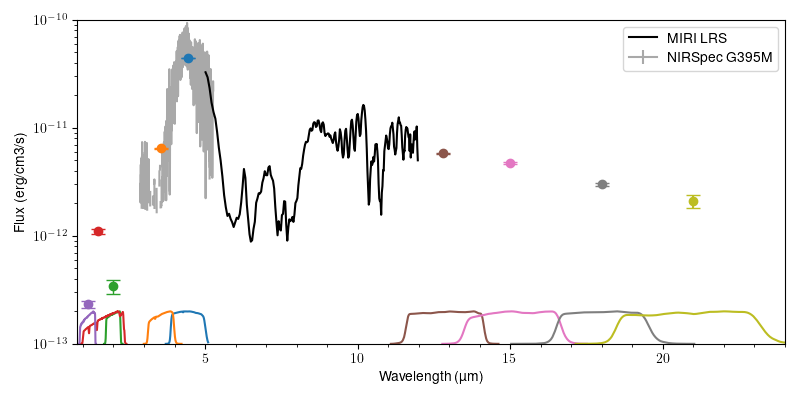}
    \caption{The combination of JWST NIRCam photometry, HST photometry, JWST NIRSpec G395M spectrum, JWST MIRI LRS spectrum, and MIRI photometry covers 91\% of the bolometric luminosity of \target{}. We also include the transmission curve of the broadband filters that share the same colors with the photometric points. The plotted photometric points from short to long wavelengths are HST/WFC3 F110W, NIRCam F150W2, NIRCam F200W, NIRCam F356W, NIRCam F444W, MIRI F1200W, MIRI F1500W, MIRI F1800W, and MIRI F2100W respectively. \rev{The MIRI photometric points are adopted from \citet{voyer2025} and the HST F110W are adopted from \cite{luhman2014}.}}
    \label{fig:fullspec}
\end{figure*}
The combination of NIRCam (1.06 -- 4.9 \um), NIRSpec (2.87-5.10 \um), and MIRI (5 -- 14 \um) data provides an extensive wavelength coverage for this cold exoplanet -- ranging from 1 to 21\um, thereby allowing us to pin down the bolometric luminosity and infer the mass and other bulk properties of \target{}.

\rev{For the bolometric luminosity calculation, we adopt the distance of 19.232 $\pm$ 0.005 pc \citep{gaiadr3}. The constructed spectral energy distribution (SED) consists of the 0.9-2.4\um\ NIRCam/F150W2 photometry, the 2.8-5.2\um\ NIRSpec/G395M spectrum, the 5.2-12\um\ MIRI/LRS spectrum, the MIRI 12.8\um, 15\um, 18\um, and 21\um\ band photometries. For flux at wavelengths longer than 23\um, which are beyond the MIRI F2100W filter's wavelength range, we used the Rayleigh-Jeans law to extrapolate the flux based on the MIRI F2100W band photometry.
We adopt a conservative estimate to the absolute flux uncertainty of 5\%\footnote{5\% is the minimum requirement for JWST instruments as listed on \url{https://jwst-docs.stsci.edu/jwst-calibration-status/jwst-absolute-flux-calibration}} for all photometric measurements. 
By integrating the spectral flux over wavelength and multiplying the photometric flux by the corresponding bandwidth, we estimated the logarithmic bolometric luminosity $\log(L/L_{sun})$ to be $-6.75 \pm 0.01$.}
The combined spectra and photometry are plotted in Figure \ref{fig:fullspec}. 
We note that the JWST observational data cover about 91\% of the bolometric luminosity for \target{}.

\rev{Based on the estimated bolometric luminosity and age, we use the Sonora Bobcat evolutionary models \citep{marley2021} to infer the mass, surface gravity, radius, and effective temperature of \target. We assume that the uncertainties in luminosity and age follow Gaussian distributions and propagate these uncertainties to the derived physical parameters using a Monte Carlo approach. Specifically, we generate 1000 samples of luminosity and age drawn from the normal distributions with means and standard deviations of $\log(L/L_{\odot}) = -6.75 \pm 0.01$ and an age of $2.0 \pm 0.5$ Gyr. Each sample is used as input to the evolutionary models. From the resulting distributions of model outputs, we infer that \target has a mass of $8 \pm 1,M_{\mathrm{Jup}}$, a surface gravity of $\log(g) = 4.26 \pm 0.08$, a radius of $1.08 \pm 0.02,R_{\mathrm{Jup}}$, and an effective temperature of $346 \pm 4$ K}
Our results showcase the superior JWST sensitivity and wavelength coverage in characterizing the fundamental mass and luminosity of one of the coldest giant exoplanets.
We discuss further the radius and the corresponding effective temperature derived from evolutionary models in Section \ref{sec:voyer2025}.

\subsection{Spectral Modeling}\label{sec:spectral_modeling}

We performed two spectral modeling approaches to fit the \target spectrum.
The first approach utilizes the model spectra grid from ``forward modeling", such as the self-consistent atmospheric models Elf-Owl \citep{mukherjee2023,mukherjee2024} and ATMOS++ \citep{leggett2023}.  
Forward modeling incorporates various physical and chemical atmospheric processes (e.g., convection, equilibrium chemistry) and solves the radiative transfer equations for an atmospheric profile that is in radiative-convective equilibrium. 
Alternatively, the second spectral modeling approach -- an atmospheric retrieval framework -- parametrizes the atmospheric temperature-pressure profile and chemical composition with many variables. 
The atmospheric retrieval framework then explores the combination of variables that best fit to the data and estimates the confidence intervals of the fitted variables with Bayesian sampling tools such as Markov Chain Monte Carlo or Nested sampling. 
These two modeling approaches provide complementary insights on the inferred atmospheric properties and on potential modeling biases.

In the following subsections, we describe common modeling parameters used in both modeling approaches, describe the two forward models Elf-Owls and ATMOS++, and introduce the first application of open-source PICASO on an atmospheric retrieval from the emission spectrum. Finally, we summarize and compare the results in Section \ref{sec:spectral_modeling}.

\subsubsection{Systematic Noise Sources, Rotational broadening, and Radial velocity}\label{sec:noise}
We included rotational broadening, radial velocity, and two systematic parameters in the two forward modeling methods (Section \ref{sec:atmos} and \ref{sec:elfowl})  as well as in the atmospheric retrieval framework (Section \ref{sec:picasor}).
We will therefore introduce here these four parameters that are used in all three spectral fitting methods.
We apply rotational broadening using the \texttt{fastRotBroad} module described by \citet{pya}. We split the data into three chunks and use the mean wavelength of each chunk to calculate the convolution kernel. 
The two systematic parameters consist of one multiplicative and one additive JWST systematic uncertainty parameter to account for the flat-field uncertainty and noise floor in our dataset.
Flat field noise is one of the major systematics for the high SNR (SNR$>$100) spectral points, with an estimated 3-10\% level. We included a multiplicative inflated noise source $\epsilon_{flat}$ to account for the flat-field noise.
In the low-SNR spectral region, imperfect background subtraction could lead to a noise floor comparable with the astrophysical signal. We therefore introduced an additive noise parameter $\epsilon_{back}$ to account for this uncertainty.
\rev{These two systematic parameters are included as part of the total uncertainties in the log-likelihood calculation (see Equation \ref{eq:loglike}).}

\subsubsection{Elf-Owl model grid}\label{sec:elfowl}
We used Elf-Owl models v2.0 \citep{mukherjee2025,wogan2025}, which provide more accurate modeling of non-equilibrium \ce{H2O}, \ce{CH4}, \ce{CO2} and \ce{CO} abundances than the Elf-Owl v1.0 model \citep{mukherjee2024}. 
The Elf-Owl models employ 7 parameters: effective temperature (\teff $\in [275, 575]$ K) ; gravity (log(g \rev{(cms$^{-2}$)})  $\in [3.23, 5.50]$); carbon-to-oxygen ratio (C/O $\in [0.5, 2.5] \mathrm{C/O_{solar}}$); eddy diffusion coefficient (log (\kzz (cm$^2$/s)) $\in [2, 9]$) ; and metallicity ([M/H] $\in [-1,1] \mathrm {M/H_{solar}} $).
The range for \teff\  was chosen to be narrower than the full range (275, 2400 K) in the  Elf-Owl models to provide higher computational efficiency.
We linearly interpolated the grid model spectra over the seven parameters.
In total there are 10 parameters including the systematics, rotational broadening, and radial velocity noted above. 
\rev{
We then obtain the best-fit model spectra by maximizing the loglikelihood function below:
\begin{align}
\log L = -0.5 \Sigma_{i=1}^N [\frac{ (M_i - D_i)^2}{\sigma_{full}^2} +2\pi \sigma_{full}^2)] \\ 
     \sigma_{full}^2 = \sigma_i^2 +(\epsilon_{flat}*Flux_{i})^2 + \epsilon_{back}^2
     \label{eq:loglike}
\end{align}
where $\sigma_i$ is the data uncertainty output by the data pipeline. We note that the multiplicative and additive uncertainties parameters ($\epsilon_{flat},\epsilon_{back}$) are applied only to the NIRSpec/G395M data and not to the NIRCam and HST/WFC3 F110W photometry. 
We adopt a uniform prior for the model parameters as shown in Table \ref{tab:forwardprior}}.
We used a Markov Chain Monte Carlo (MCMC) method with emcee \citep{foreman-mackey2013} to sample the posterior distribution of \rev{the fitted parameters.}
We used 1000 walkers running for 20,000 steps.
As suggested in the emcee package, we confirmed that the MCMC results had converged as the number of steps is more than 50 times longer than the maximum autocorrelation length (157) of the parameters.

\subsubsection{ATMOS++ grid models}\label{sec:atmos}

The spectral fitting routine using the ATMOS++ model \citep{atmos++} consisted of eight parameters, including effective temperature, surface gravity $\log(g)$,radius, metallicity [M/H], rotational broadening vsini, two noise parameters ($\epsilon_{back}$ and $\epsilon_{flat}$), and radial velocity.
Similar to the spectral fitting routine of the Elf-Owl model, we used emcee to obtain the parameters with maximal loglikelihood and sampled the posterior distribution of the parameters.
For ATMOS++ models, our emcee employed 100 walkers and ran for 10,000 steps. The ATMOS++ MCMC sampling process converged with the maximum autocorrelation length of 105 to be at least 50 times smaller than the 10,000 steps.

\begin{table}[ht]
\centering
\caption{Model parameters and prior ranges for ELF--OWL and ATMO(S)++ atmospheric retrievals. All priors are uniform unless otherwise stated.}
\label{tab:forwardprior}
\begin{tabular}{lcc}
\hline
\hline
\textbf{Parameter} 
& \textbf{ELF--OWL} 
& \textbf{ATMOS++} \\
\hline
 $T_{\mathrm{eff}}$ (K) 
& $\mathcal{U}(275,\,575)$ 
& $\mathcal{U}(250,\,1200)$ \\

 $\log g$ (cgs) 
& $\mathcal{U}(3.23,\,5.50)$ 
& $\mathcal{U}(2.5,\,5.5)$ \\

 $\log K_{zz}$ 
& $\mathcal{U}(2,\,9)$ 
& -- \\

 [M/H] ([M/H]$_{\odot}$=0)
& $\mathcal{U}(-1.0,\,1.0)$ 
& $\mathcal{U}(-0.5,\,0.5)$ \\

C/O ratio 
& $\mathcal{U}(0.5,\,2.5)$ 
& -- \\

 Radius ($R_{\mathrm{Jup}}$) 
& $\mathcal{U}(0.5,\,2.0)$ 
& $\mathcal{U}(0.5,\,2.0)$ \\

 $v\sin i$ (km s$^{-1}$) 
& $\mathcal{U}(0.1,\,200)$ 
& $\mathcal{U}(0,\,200)$ \\

  RV (km s$^{-1}$) 
& $\mathcal{U}(-100,\,100)$ 
& $\mathcal{U}(-100,\,100)$ \\

 $\log \epsilon_{\mathrm{flat}}$ 
& $\mathcal{U}(-4,\,0)$ 
& $\mathcal{U}(-4,\,0)$ \\

 $\log \epsilon_{\mathrm{back}}  (cgs)$ 
& $\mathcal{U}(-14,-11)$ 
& $\mathcal{U}(-14,-11.3)$ \\
\hline
\hline
\end{tabular}
\end{table}

\subsubsection{The atmospheric retrieval framework with PICASO}\label{sec:picasor}
In our atmospheric retrieval, we used PICASO \citep{batalha2019} for the radiative transfer calculation and Pymultinest \citep{buchner2014} for sampling the posterior distribution.
In our retrieval, we setup a log-uniform pressure grid from 1mbar to 100bars with 55 layers.
We included altitude-independent volume mixing ratios of \rev{seven} molecules: \ce{CH4}, \ce{H2O}, \ce{CO2}, \ce{H2S}, \ce{PH3}, and \ce{NH3}, \rev{\ce{CO}} in our retrieval.
\rev{We also include \ce{H2-He} and \ce{H2-H2} collision-induced absorption (CIA) opacity.}
Based on the contribution function of the best-fit Elf-Owl grid models, more than 99.5\% of the 1--5.2\um top-of-atmosphere flux is emitted from the 0.35--17 bar pressure regime.
Informed by the contribution function, we placed eight  temperature-pressure (TP) ``knots'' that were used for spline interpolation in the parameterization of the TP profile.
We placed five TP knots in the 0.35--17\,bar photosphere regime, with two additional TP knots above the photosphere separated by 1dex pressures, and one TP knot 1dex below the photosphere. In total, there are therefore eight TP knots spanning over 0.0035 - 86  bar.
Because of the weak irradiation from the white dwarf at a projected distance of 2500 au, we assume there was no thermal inversion and limit the temperature to monotonically increase with higher pressure.
We assume a cloud with a grey opacity (i.e., a zero single scattering albedo and a zero asymmetry parameter)  and model the cloud opacity exponentially decreasing with lower logarithmic pressure:
\begin{equation}
    \tau_{cloud}(p) = \tau_{base}*e^{f_{sed}[\log(p) - log(p_{base})]}
\end{equation}
where $\tau_{base}$ is the cloud opacity at cloud base pressure $p_{base}$ and $f_{sed}$ is a parameter that controls the steepness of the exponential decay of cloud opacity. The three parameters -- $\tau_{base}$, $p_{base}$,$f_{sed}$ -- are used to characterize the possible cloud opacity in the retrieval framework.


We setup the prior ranges of temperature-pressure profiles motivated by the forward modeling results and grid models. We first normalized the TP profiles with effective temperatures ranging from 300 to 400\,K such that they have the same temperature at the photosphere ($\sim$2.5bar). These normalized TP profiles reflect the possible TP profiles over various gravity, non-equilibrium chemistry, metallicity, and C/O ratios explored in the Elf-Owl models. 
Starting from the temperature knot T4(P=1bar), we then assumed the T4 prior range to be $\pm$100K of the best-fit grid model value, and further multiply the boundary value by $\pm$40\% as the prior range of the T4 knot. Given the prior range of T4, we then rescaled the normalized TP profiles to match the T4 values. 
The minimum and maximum values of the scaled TP profiles were then used as the prior range for the rest of the temperature knots.
We list the prior ranges for the other parameters in Table \ref{table:prior}.
\rev{The log-likelihood equation is the same that used in forward modeling (i.e., Equation \ref{eq:loglike}).}

Because it is common to get a retrieval framework bias toward a high-gravity solution that is inconsistent with evolutionary models \citep[e.g.,][]{zalesky2019,kitzmann2020}, we introduced a Gaussian prior for the surface gravity informed by the age and the measured luminosity as described in Section \ref{sec:lbol}.
Based on the inferred surface gravity, the Guassian prior for surface gravity $\log(g)$ is a normal distribution with a mean of 4.26 and a standard deviation of 0.08.
In total, there are three retrieval runs: (1) Cloudless retrieval with a uniform gravity prior, (2) Cloudless retrieval with a Gaussian gravity prior, and (3) Cloudy retrieval with a Gaussian gravity prior.
 We used the nested sampling package Pymultinest \citep{buchner2014} with 2000 live points, constant efficiency argument set to False, and importance sampling argument set to True to sample the posterior distribution. 
 We list the retrieval results in Table \ref{table:prior} and compare the key retrieval results to the forward models in Table \ref{tab:results}.

\subsection{Spectral Modeling Results}\label{sec:spectral_modeling}

\rev{\begin{figure*}
\includegraphics[width=1.0\linewidth]{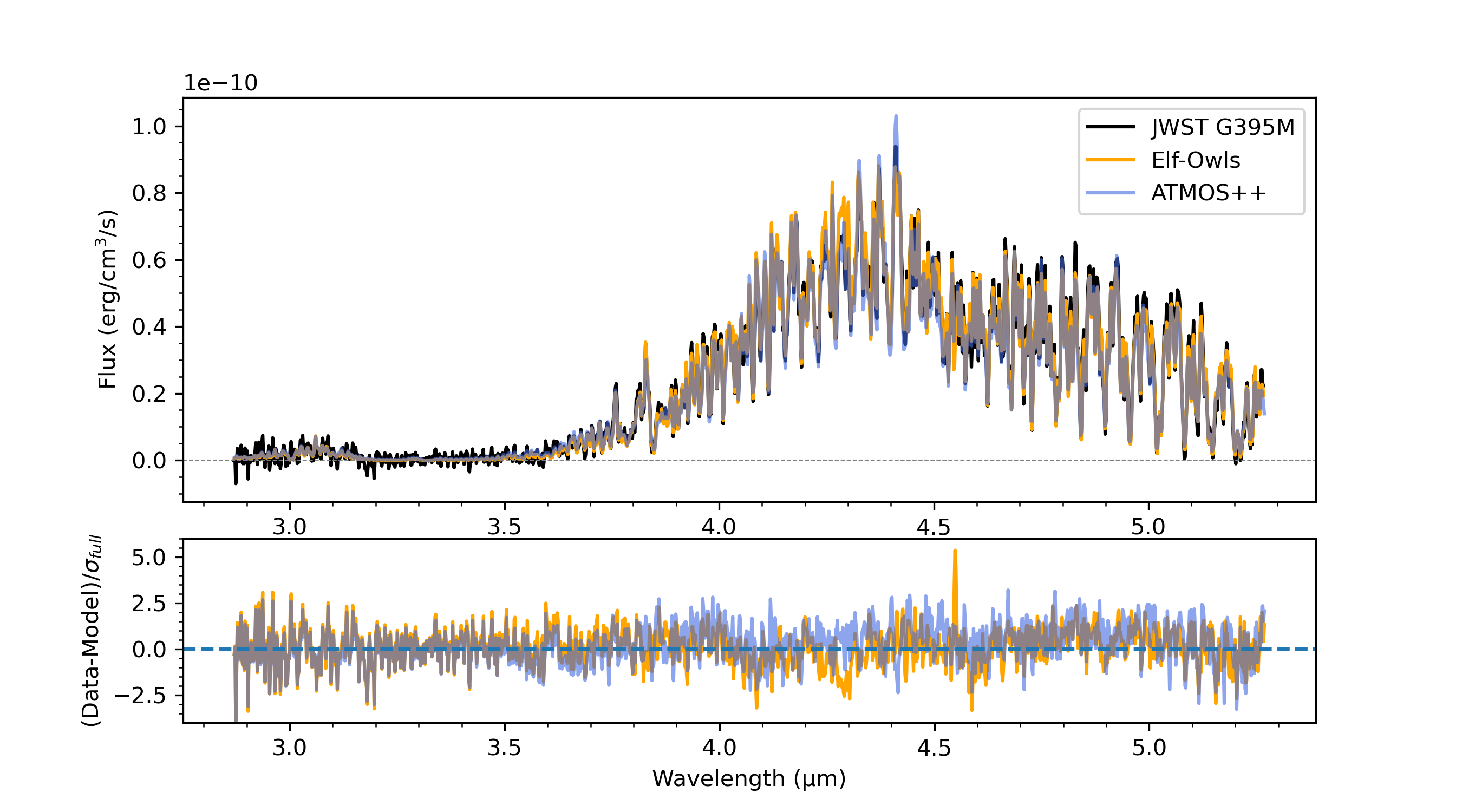}
\caption{Upper panel: the best-fit Elf-Owl (orange line) and ATMOS++ (blue line) model spectra compared to the JWST G395M spectrum (black line). Bottom panel: The residuals between the models and data normalized by the total uncertainties. The sharp positive residual feature at 4.55\,\um\ of Elf-Owl model arises from the model \ce{CH4} opacity line list that includes \ce{CH3D} at solar mixing ratio.}
\label{fig:forwardmodels}
\end{figure*}}

\begin{figure*}
    \centering
    \includegraphics[width=.52\linewidth]{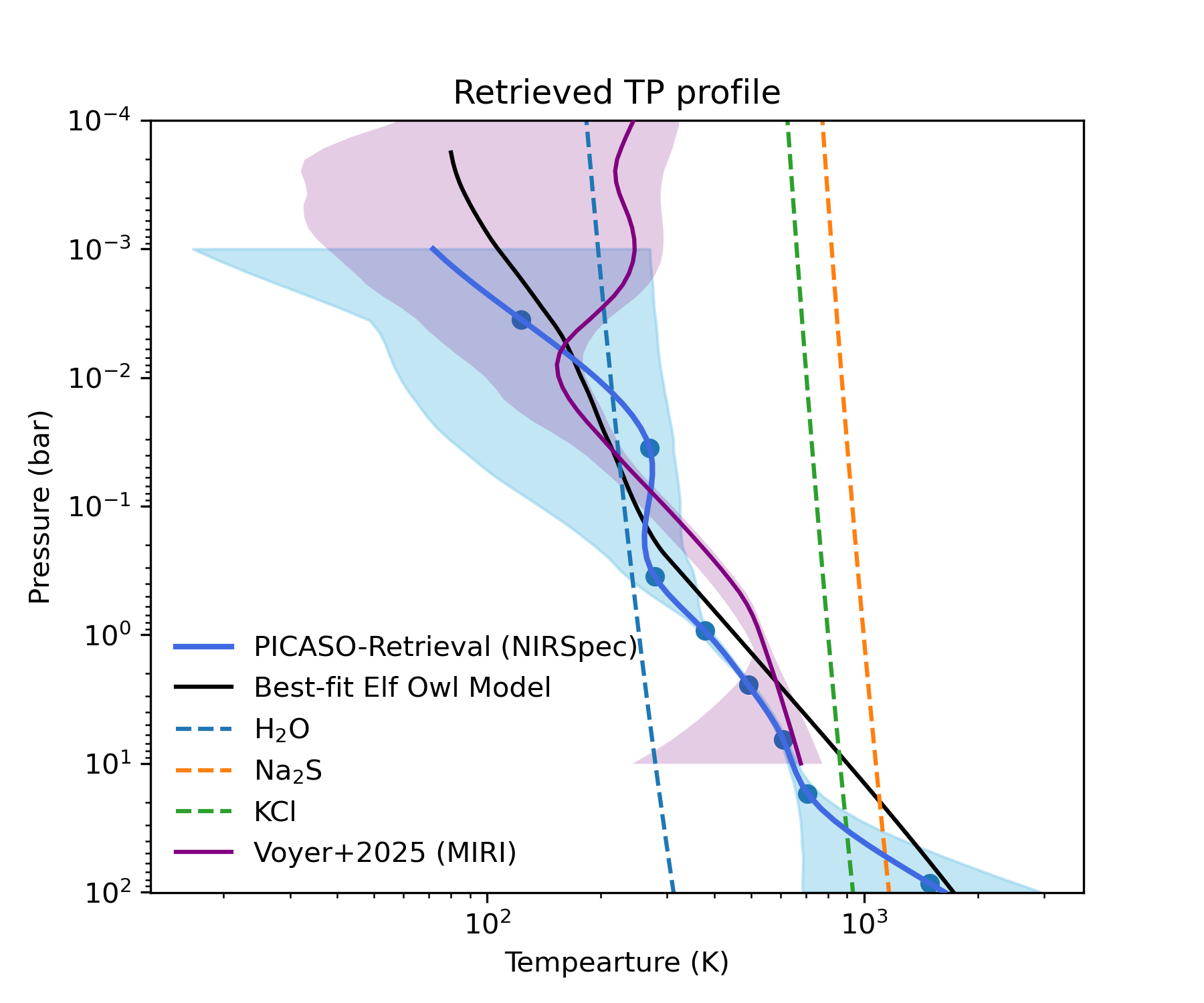}%
\includegraphics[width=.58\linewidth]{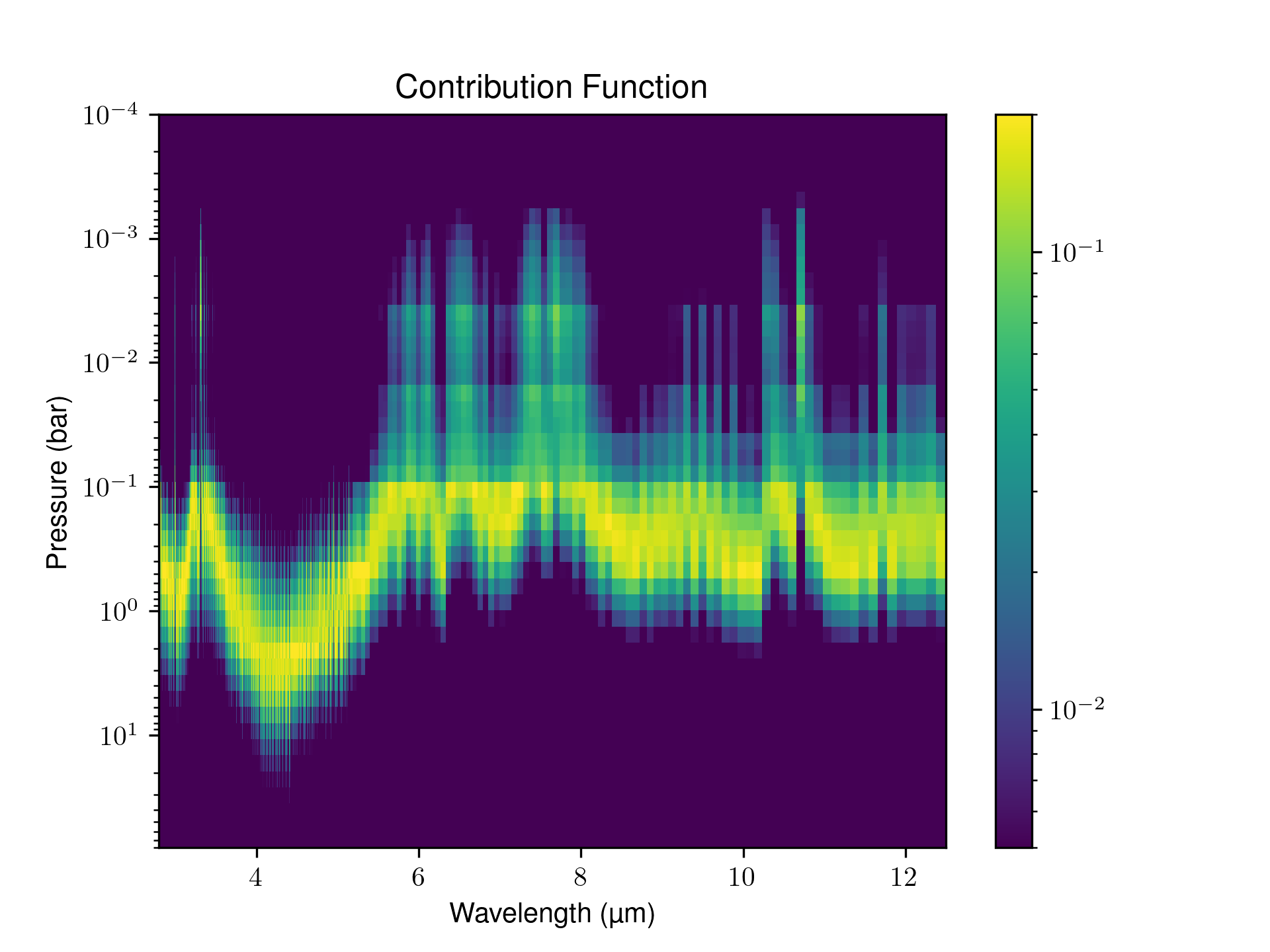}
    \caption{\textbf{Left panel:} The best-fit PICASO retrieval temperature-pressure (TP) profile is plotted as a solid blue line with the temperature-pressure knots indicated in solid blue circles. The light blue shaded region indicates the 3$\sigma$ confidence interval of the retrieved TP profiles. The tightly constrained TP profiles in the 1--10 bar range match with the peak of the 3--5\um contribution function plotted in the right panel.
    The retrieved TP profile based on MIRI-LRS 5--11.7\um spectrum in \citet{voyer2025} is plotted in solid purple along with the 3$\sigma$ uncertainty plotted as light-purple shaded region.
    The condensation curves of \ce{H2O}, \ce{Na2S}, and \ce{KCl} at solar metallicity are plotted with blue, orange, and green dashed lines respectively. 
    For reference, the solid black line shows the TP profile of the best-fit Elf Owl grid model ([M/H] = -1, C/O = 1.5, log(g) = 3.23, log(\kzz) = 7, \teff = 350\,K). \textbf{Right panel:} The contribution function map of the best-fit PICASO retrieval results shows that most of the 3--5\um flux is emitted from 0.1--10 bar region.}
    \label{fig:tp}
\end{figure*}

\begin{table*}[ht]
\centering
\begin{tabular}{lccccc}
\hline
\textbf{Parameter} &\textbf{PICASO Retrieval}    & \textbf{PICASO Retrieval} & \textbf{Elf-Owl v2.0}  & \textbf{ATMOS++} & {\textbf{PICASO Retrieval w/}} \\ 
& cloudy, $\mathcal{N}$[log(g)]& cloudless,$\mathcal{N}$[log(g)]&&& { cloudless, $\mathcal{U}$[log(g)] }  \\ \hline
Teff             & $386 \pm 4$ & $387\pm 4$ & $358^{+3}_{-3}$     & 402 $\pm1$                      &                            395 $\pm$ 4 \\
logg             &4.42$\pm{0.04}$ & $4.43^{+0.04}_{-0.05}$ &  $3.32^{+0.03}_{-0.03}$    & 4.50 $\pm0.01$                       &  $4.9 \pm 0.1$                           \\
Radius           & 0.83 $\pm 0.02$ & $0.82\pm0.02$   & $0.83^{+0.01}_{-0.01}$     & 0.82$\pm0.01$                        &  $0.78 \pm 0.01$                           \\ 
{[M/H]}           & -0.25$\pm 0.04$ (A21)  &$-0.25\pm0.04$ (A21)   & $-0.93^{+0.03}_{-0.04}$ (L09)   & 0.02$\pm0.01$ (A09)                        &   +0.14 $\pm$ 0.1 (A21) \\
Absolute C/O       & $0.75^{+0.03}_{-0.02}$  & $0.76 \pm0.03$         &   $0.59 \pm 0.02$ (bulk)                   & 0.55 (fixed) &   $0.79\pm0.02$ \\
v$\sin$(i) ($\mathrm{km s^{-1}}$)   & $<28$         & $<28 $      & $<21$ & $<26$                             &  $<20$                        \\ 
RV ($\mathrm{km s^{-1}}$) &     $32 \pm 1$                & 32$\pm$1  & $32^{+1}_{-1}$ & 37$\pm2$                         & 32$\pm$0.7                          \\ 
log($\epsilon_{flat}$)    & $-1.48 \pm 0.03$     &$-1.48\pm0.03$        & $-1.19^{+0.02}_{-0.02}$ & -1.15$\pm0.02$                       & $-1.50\pm0.02$                            \\ 
log($\epsilon_{back}$)  & $-11.80 \pm 0.01.$ &$-11.81\pm0.01$     &  $-11.81^{+0.01}_{-0.01}$          & -11.75$\pm0.01$                      &  $-11.80\pm0.01$                       \\ 
logKzz &-&-& $7.2^{+0.3}_{-0.2}$     & - & - \\ 
$\chi_{r}^2(\sigma_{data})$      & 26.9   &27.5&30&76&27.9 \\
Num of params &24&21 &10 &8&21 \\
$\Delta$BIC &0 & -80 & +306 & +161 &-85    \\ \hline
\end{tabular}
\caption{Comparison of different models (Elf-Owl, ATMOS++, PICASO Retrieval) for key parameters. Either a gaussian ($\mathcal{N}$) or uniform ($\mathcal{U}$) range is used when sampling the prior of surface gravity parameter log(g) in PIACSO-R. A21, L09, A09 refer to the solar metallicities reported in \citet{asplund2021}, \citet{Lodders2009}, and \citet{asplund2009} correspondingly. In the Elf-Owl models, the absolute C/O ratio refers to bulk C/O that includes oxygen in silicate clouds and in molecular gases. }\label{tab:results}
\end{table*}

We present the key fitted parameters of the best-fit models in Table \ref{tab:results}.
\rev{The best-fit PICASO retrieval spectrum is plotted in Figure \ref{fig:spec} while the best-fit Elf-Owl and ATMOS++ model spectra are plotted in Figure \ref{fig:forwardmodels}.
In Figure \ref{fig:spec}, we interpret that the negative residuals in the 4-4.2 \um\, are likely caused by the retrieval profile temperature-pressure profile being too hot and/or the methane opacity is underestimated  around 2-10 bars, the pressure region at which the contribution function in 4-4.2 \um\ peaks in Figure \ref{fig:tp}.}
The retrieval results (i.e., the first column) have lower reduced chi-squared and Bayesian Information Criterion (BIC) values than the Elf-Owl and ATMOS++  models.
Both the cloudless and cloudy retrieval yield similar results.
While the cloudy retrieval has a higher BIC value than the cloudless retrieval, we chose to adopt the maximal likelihood solutions of cloudy retrieval as our best-fit solution to minimize potential biases due to cloud opacity. 
We note that the difference between cloudless retrieval results with a uniform and with a Gaussian prior indicates that the retrieval results are sensitive to the prior for surface gravity, which are set by the age and measured bolometric luminosity using evolutionary models.

The best-fit TP profile of PICASO retrieval is plotted in the left panel of Figure \ref{fig:tp}.
The effective temperature of PICASO retrieval is 386$\pm$4 K, in agreement with the ATMOS++ and Elf-Owls' fitted temperatures within 30\,K.
We also include the condensation curves of \ce{H2O}, \ce{Na2S}, and \ce{KCl} in Figure \ref{fig:tp} for reference. 
\rev{In the 1-10 bar region where the TP profile is well constrained, the condensation curves of these three species do not intersect with the retrieved TP profile. Therefore, the retrieval results are consistent with the prediction of the condensation curves that no significant cloud opacity is present in the 1-10 bar region.}
The retrieved mass is $7\pm1$ M$_{Jup}$, consistent with the mass that is estimated based on the observed bolometric luminosity in Section \ref{sec:lbol}.
Except for the ATMOS++ which has a fixed solar C/O ratio, both PICASO Retrieval and Elf-Owl models prefer above solar C/O ratios with subsolar metallicities.
All three models provide a similar radius of around 0.82 \rjup that is smaller than the evolutionary model's prediction of 1.1 \rjup. We refer further discussion on the radius to Section \ref{sec:radius}. 
We note that all three modeling approaches provide similar systematics parameters including $\epsilon_{back}$, $\epsilon_{flat}$, RV, and $v \sin$(i). 
The fitted $\epsilon_{flat}$ is about $3\%$, which is consistent with the published absolute flux uncertainty of $\sim 3\%$ for the NIRSpec IFU data.\footnote{\url{https://jwst-docs.stsci.edu/jwst-calibration-status/nirspec-calibration-status/nirspec-ifu-calibration-status}}.  

\begin{table*}[ht]
\centering
\begin{tabular}{lcc}
\hline
\textbf{Parameter}      & \textbf{Prior} & \textbf{PICASO cloudy retrieval} \\ \hline
$\log(g [\mathrm{cms^{-2}}]$)                  & $\mathcal{N}$($\mu=$4.26, $\sigma=$0.08)  & 4.42 $\pm 0.04$         \\
log(VMM(CO))                     & U(-13, -2)  & -6.31 $\pm 0.04$       \\
log(VMM(PH$_3$))                 & U(-13, -2)       &-11.1$^{+1.3}_{\red{-1.2}}$\\
log(VMM(CH$_4$))                 & U(-10, -1)   & -3.51 $\pm 0.04$     \\
log(VMM(H$_2$O))                 & U(-10, -1)       & -3.39 $\pm 0.04$  \\
log(VMM(CO$_2$))                 & U(-13, -2)      &-8.95$^{0.05}_{-0.06}$   \\
log(VMM(H$_2$S))                & U(-13, -2)     & -4.74$^{+0.06}_{-0.07}$    \\
log(VMM(NH$_3$))                 & U(-13, -2) &-4.58$^{+0.04}_{-0.05}$        \\
T$_1 (P=3.5\,$mbar)            & U(48, 402) K  & 113$^{+67}_{-44}$ K  \\
T$_2 (P=35\,$mbar)            & U(84, 537) K   & 207$^{+50}_{-67}$ K \\
T$_3 (P=0.35\,$bar)            & U(164, 690) K  & 281$^{+23}_{-18}$ K \\
T$_4 (P=0.93\,$bar)            & U(216, 772) K  & 378$\pm 3$ K  \\
T$_5 (P=2.45\,$bar)            & U(300, 888) K  & 493$^{+4}_{-2}$ K   \\
T$_6 (P=6.48\,$bar)            & U(378, 1168) K & 608$\pm 4$ K   \\
T$_7 (P=17.1 \,$bar)            & U(470, 1564) K & 710$^{+16}_{-17}$ K   \\
T$_8 (P=85.4\,$bar)            & U(649, 2526) K  &  $1800^{+483}_{-605}$ K \\
$\log(P_{\rm cloud-base} (bar)$ &  U(-2.8,1.8) &  (unbounded)\\
$\log(\tau_{\rm cloud-base})$ & U(-5,2) & (unbounded)\\
\fsed & U(0.1,80) & (unbounded) \\
Radius ($R_{\textrm jup}$)               & U(0.5, 2.0)   & 0.82$\pm 0.02$        \\
v$\sin$ i (km/s)          & U(1$\times$10$^{-3}$, 200)  & ${<28\, (3\sigma)}$\\
Radial Velocity (km/s)   & U(-100, 100)    & 32 $\pm$ 1      \\
noise floor $\log (\epsilon_{\textrm flat})$                 & U(-4, 0)    & -1.48$\pm 0.03$         \\
flat-field uncertainty $\log (\epsilon_{\textrm back})$            & U(-14, -11)  & -11.8$\pm 0.01$     \\ \hline
\end{tabular}
\caption{The priors and median values of the marginalized posterior parameters of the age-constrained cloudy PICASO retrieval framework. The +/- 1-$\sigma$ uncertainties are shown along with the median values. The T$_i$ represent the pressure-temperature knots. The numbers in red of the uncertainties \ce{PH3} and vsin(i) indicate that the lower-bound uncertainties are not well constrained. See Section \ref{sec:noise} for the definition of the noise floor and flat-field uncertainties. }\label{table:prior}
\end{table*}

\subsection{Elemental Abundances Ratio}\label{sec:elementratio}
Based on the PICASO retrieval results, we derived the elemental abundance ratios of C, N, O, S, and P. In the retrieval, we assumed that molecular hydrogen and helium make up of the background gas in addition to the included six molecules, i.e,. $\Sigma^6_{i=1} X_i$ + X$_{H_2}$ + X$_{He}$ = 1. The ratio between the volume mixing ratio of molecular hydrogen and helium was assumed to be 0.8375/0.1625 or 5.153. For each molecule i, we used the median values of the retrieved volume mixing ratio X$_i$ and used the corresponding 16- and 84 percentile values of the marginalized posterior distribution to calculate the 1-$\sigma$ uncertainty. 
We calculated the elemental abundance ratios as follows:
\begin{align}
\text{C/H} &= \frac{X_{\ce{CH4}} + X_{\ce{CO}} + X_{\ce{CO2}}}{2X_{\ce{H2}}}, \\
\text{N/H} &= \frac{X_{\ce{NH3}}}{2X_{\ce{H2}}}, \\
\text{O/H} &= \frac{X_{\ce{H2O}} + X_{\ce{CO}} + 2  X_{\ce{CO2}}}{2X_{\ce{H2}}}, \\
\text{S/H} &= \frac{X_{\ce{H2S}}}{2X_{\ce{H2}}}, \\
\text{P/H} &= \frac{X_{\ce{PH3}}}{2X_{\ce{H2}}}.
\end{align}
The results were then normalized by the solar abundance of \citet{asplund2021} and are presented in Table \ref{tab:element_ratios}. \rev{The P/H is a 3$\sigma$ upper limit because the marginalized posterior distribution of PH3 abundance is unbounded toward lower values.}. We also derived the C/O and C/S ratios using the calculated C/H, O/H, and S/H.
We conclude that all the derived abundance ratios except S/H are significantly subsolar, while the S/H ratio is about 2-$\sigma$ below the solar abundance. 
The C/O ratio of 0.76$\pm 0.03$ is significantly above solar value of about 0.59 while the C/S ratio of 17$^{+3}_{-2}$ is about 2-$\sigma$ below the solar C/S value of 22.

We further discuss the interpretation of the subsolar elemental abundance ratios in Section \ref{sec:subsolar}.

\begin{table}[h]
    \centering
    \begin{tabular}{cc}
        \hline 
\shortstack{} &\shortstack{Elemental Abundance Relative\\to \citet{asplund2021}} \\
        \hline
        C/H  &  $0.56 \pm 0.05$ ($1.85^{+0.16}_{-0.17}\times 10^{-4}$) \\
        N/H  & $0.21 \pm 0.02 $ ($1.60^{+0.15}_{-0.16}\times 10^{-5}$)\\
        O/H  & $0.44 \pm 0.04$  ($2.47^{+0.23}_{-0.23}\times 10^{-4}$)\\
        S/H  & $0.77^{+0.11}_{-0.12}$ ($1.11^{+0.18}_{-0.17}\times 10^{-5}$)\\
        P/H & $<$ $2.5\times 10^{-3}$\ ($<7.4\times10^{-10}$,3$\sigma$) \\
         C/O  & 1.28 $\pm$ 0.05 ($0.76 $ $\pm$0.03) \\
         C/S & 0.73 $\pm$ 0.1 ($17 ^{+3}_{-2}$) \\
        \hline
    \end{tabular}
    \caption{The derived relative elemental abundance ratio based on the PICASO retrieval results. The solar values are adopted from \citet{asplund2021}. The values in the brackets are absolute values and the 3$\sigma$ limit for P/H.}
    \label{tab:element_ratios}
\end{table}

 \subsection{A New Framework for Deriving \kzz\ at Multiple Quench Levels}\label{sec:kzz}
\begin{figure*}
    \centering
    \includegraphics[width=1.0\linewidth]{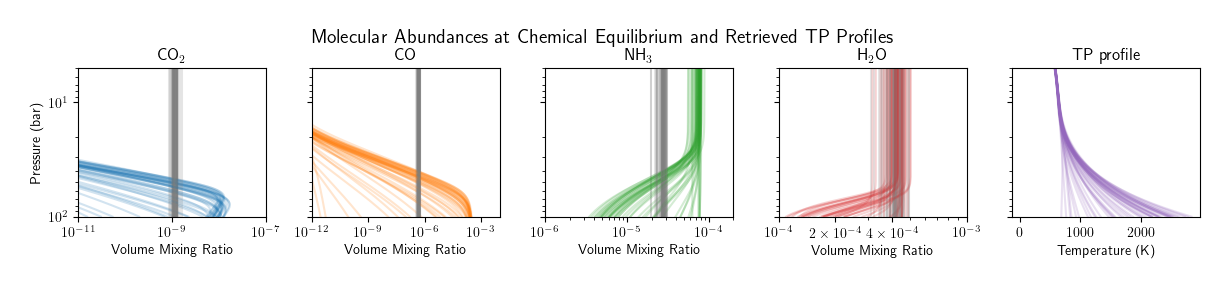}
    \includegraphics[width=1.0\linewidth]{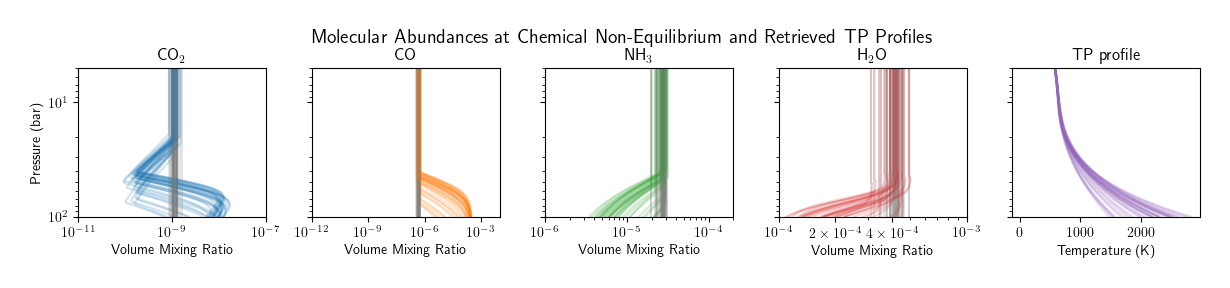}
    \caption{\textbf{Top row}: Output from the chemical equilibrium models described in Section \ref{sec:kzzmethod}. Only 50 samples are plotted for clarity in this figure. Volume mixing ratios of \ce{CO2}, CO, \ce{NH3}, and \ce{H2O} are shown as blue, orange, green, and red lines, respectively, based on 50 posterior samples from the PICASO retrievals. Grey vertical lines in the first four panels represent the retrieved mixing ratios for each molecule. The corresponding temperature-pressure (TP) profiles are plotted as purple lines in the fifth panel. By identifying where the retrieved abundances (grey lines) intersect with the equilibrium curves (colored lines), we estimate the quench pressures.
    \textbf{Bottom row}: Same sampled profiles as that the top row,  except we have quenched each abundance profile where it intersects the vertically-constant posterior abundance at quenched pressure $P_{quench}$. 
    The samples with retrieved abundances greater than chemical equilibrium abundances (e.g., CO) are excluded in bottom row because the quenched pressures cannot be estimated in these cases. Comparing the TP profiles between top and bottom rows, the excluded samples mostly have cold TP profiles and low CO chemical equilibrium abundances. The sharp inversion of the \ce{CO2} abundance near 50 bars is equivalent to the quenched CO pressure. See Section \ref{sec:kzzmethod} for other details of the non-equilibrium models.
    }
    \label{fig:chemeq}
\end{figure*}

With the retrieved abundances of species like \ce{CO}, \ce{CO2}, \ce{CH4}, and \ce{NH3}, we can test if the abundances deviate from the chemical equilibrium values.
If the abundances are out of chemical equilibrium, we can then investigate the dynamical timescale needed to drive the abundances away from the equilibrium state, and hence estimate the equivalent eddy diffusion coefficient \kzz\.
In the following Section \ref{sec:kzzmethod}, we first describe the methods used to calculate the chemical equilibrium and disequilibrium abundances, and the corresponding eddy diffusion coefficients for each species. We then present and discuss the results in Section \ref{sec:kzzresults}.

\subsubsection{Chemical Equilibrium and Non-equilibrium models} \label{sec:kzzmethod}

We used the equilibrium chemistry solver in Photochem v0.6.5 \citep{wogan2024} to compute molecular abundances for the temperature-pressure profiles, C/O, and metallicities derived from the PICASO retrieval posterior samples.
We adopted the elemental abundance of \citet{asplund2021}, which has a solar carbon-to-oxygen ratio of 0.59.
Similar to the description in Section \ref{sec:elementratio}, we approximated the bulk metallicity [M/H] with the retrieved C/H.
We modeled non-solar C/O by decreasing the elemental oxygen abundance from the solar value, analogous to the sequestration of oxygen into silicate clouds at depth, which can increase the upper atmospheric C/O ratio to a level that is higher than the bulk C/O ratio.
The output of chemical equilibrium models are plotted in Figure \ref{fig:chemeq}.

In the chemical non-equilibrium models, we introduced a physical parameter ``quenched pressure" to identify the pressure region in which a species ceases to be in chemical equilibrium.
We assume that the quenched pressures of \ce{CO}, \ce{CO2}, and \ce{NH3} are deeper than the photospheric pressures (i.e., $P_{quenched} > P(\tau=1)$, which is shown by the contribution function in Figure \ref{fig:tp}), so that the all three molecular abundances are quenched at the photosphere.
Under this assumption, we then estimated the quenched pressure by solving for the pressure at which the chemical equilibrium abundances are equal to the PICASO retrieved abundances of \ce{CO}, \ce{CO2}, \ce{H2O} and \ce{NH3}.
As a result, the molecular abundances in non-equilibrium models are vertically constant and equal to the retrieved abundances until they reach their quenched pressures, as shown in the bottom row of Figure \ref{fig:chemeq}.

Under a chemical equilibrium state, CO decreases while \ce{NH3} increases with lower pressure and temperature. Their abundances become uniform for P$<P_{quenched}$.
In the case of \ce{CO2}, the abundance profile exhibits a distinct inflection point where the abundance increases with lower pressure. 
As discussed in \citet{beiler2024, mukherjee2025,wogan2025}, this behavior corresponds to a regime in which \ce{CO2} remains in chemical equilibrium with \ce{CO}, while \ce{CO} itself is out of equilibrium.
The \ce{CO2} abundance increases with lower pressure until it is quenched and becomes uniform with pressure. 
We note that the \ce{H2O} abundances are almost the same in both chemical equilibrium and non-equilibrium models.
Both chemical equilibrium and non-equilibrium models show similar \ce{H2O} abundances because, under CO quenching, the amount of \ce{H2O} consumed to produce \ce{CO} from \ce{CH4} is small compared to the total \ce{H2O} abundance.

After estimating the quenched pressures, we then estimated the chemical timescales for each molecule at the quenched pressures. 
We used Equations 12, 13, 14, 44, and 32 in \citet{zahnle2014} to calculate the chemical timescale ($\tau_{chem}$) of \ce{CO}, \ce{CO2}, and \ce{NH3}:

\begin{align}
t_{\rm q1}^{\rm }(\ce{CO}) & = 1.5 \times 10^{-6} p^{-1} m^{-0.7} \exp\left( \frac{42000}{T} \right)s \\
t_{\rm q2}^{\rm}(\ce{CO}) &= 40 p^{-1} \exp\left( \frac{25000}{T} \right)s \\
\tau_{\rm chem}(\ce{CO}) &= \left( \frac{1}{t_{\rm q1}} + \frac{1}{t_{\rm q2}} \right)^{-1}s\\
\tau_{\rm chem}(\ce{CO_2}) &= 1.0{\times}10^{-10} p^{-0.5} \exp\left( \frac{38000}{T} \right)s \\
\tau_{\rm chem}(\ce{NH_3}) &= 1.0{\times}10^{-7} p^{-1} \exp\left( \frac{52000}{T} \right)s      \label{eq:kzz} 
\end{align}

Finally, we assumed that the mixing length scale $L$ is the same as the pressure scale height ($H$)\footnote{See also \citet{Bordwell2018} for derivation of mixing length scale that is chemical-rate dependent.} and derive the \kzz\ value:
\begin{equation}
        K_{zz} (X) = \frac{L^2}{\tau_{chem}(X)} = \frac{H^2}{\tau_{chem}(X)}\label{eq:kzz2}
\end{equation}
We discarded samples of those TP profiles that are too cold such that the equilibrium abundances \ce{CO2} and \ce{CO} are lower than the retrieved values even at 100 bars, the maximum pressure limit of our model grid. 
Therefore, there are fewer samples in the bottom panel than in the top panel in Figure \ref{fig:kzz}.

\subsubsection{Derived Eddy Diffusion Coefficients}\label{sec:kzzresults}

To our knowledge, this is the first study that leverages observational data to derive altitude-dependent \kzz\ values via their quenched pressures. 
By investigating \kzz\ at different quenched pressures in an atmosphere, our aim is to establish this as a proof-of-concept and understand: 1) what \kzz\ values this methodology produces, 2) if these \kzz\ values appear to be physically motivated, 3) to what degree they are consistent amongst themselves, and 4) how are these \kzz\ compared to self-consistent models.

\begin{figure*}
    \centering
    \includegraphics[width=1.0\linewidth]{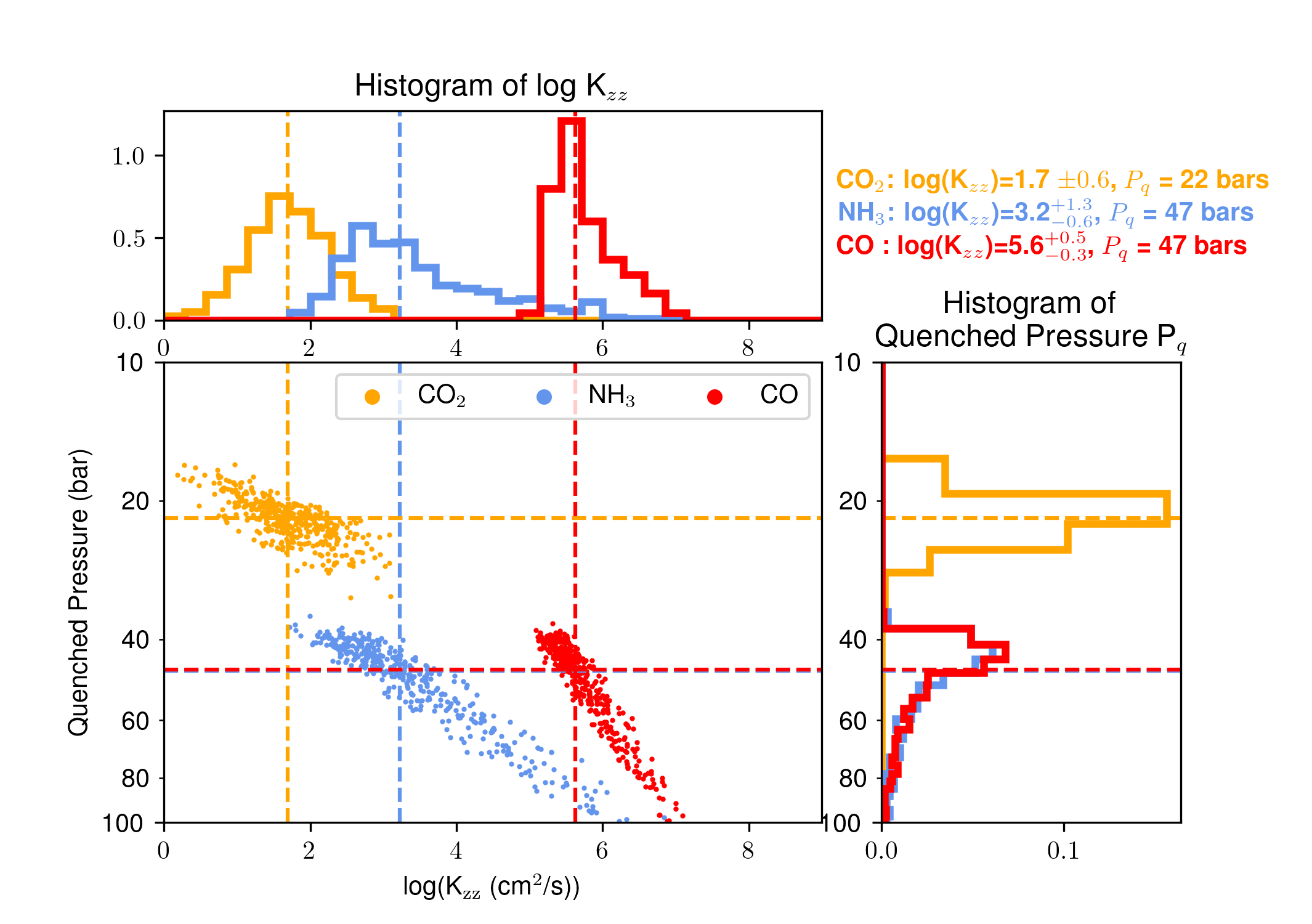}
    \caption{
    \textbf{Bottom Left}: the derived \logkzz\ of \ce{CO}, \ce{CH4}, and \ce{NH3} as a function of quenched pressures based on the temperature-pressure profile and abundances from the 500 posterior samples of the PICASO retrieval.
    The median values of \logkzz\ and quenched pressures of are plotted in colored dashed lines.
    \ce{CO2} has the lowest median \kzz\ among the 3 species and has the lowest quenched pressure (22 bars) while \logkzz\ of CO is the highest with a median quenched pressure of 47 bars. The derived \logkzz\ of the three molecules increase with higher quenched pressures. \textbf{Top left and bottom right:} the histograms of the derived \kzz\ and quenches pressures are plotted in the top and right panel respectively.  See text in Section \ref{sec:kzzresults} for details about the derivation of \kzz. }
    \label{fig:kzz}
\end{figure*}

Based on the method described in the previous section, we randomly sampled 500 posterior samples of the retrieval results to derive the \kzz\ values and quenched pressures for the \ce{NH3}, \ce{CO}, and \ce{CO2} species.
The quenched pressures of \ce{CO}, \ce{NH3}, and \ce{CO2} are 47$_{-7}^{+16}$, 47$_{-7}^{+19}$, and $22\pm 2$ bars respectively with the corresponding $\log$(\kzz) values of $1.7\pm 0.6$, 3.2 $^{+1.3}_{-0.6}$, and 5.6$^{+0.5}_{-0.3}$ respectively.
The quenched pressure and \kzz\ in Figure \ref{fig:kzz} are positively correlated because \kzz\ is inversely proportional to the chemical timescale (see Eq. \ref{eq:kzz}), which decreases with higher pressure and temperature.

The derived quenched pressures of \ce{NH3} and \ce{CO} are strikingly similar, both are around $\sim$50 bars, but the median derived \logkzz\ differ by more than two orders of magnitude, or differ from each other at 1.9$\sigma$.
Based on the posterior distributions, the \kzz\ of \ce{CO2} is 1.7$\sigma$ lower than that of \ce{NH3} and 5.8$\sigma$ lower than that of \ce{CO}.
Therefore, our results suggest that \kzz\ decreases with lower pressure based on the \ce{CO2} and \ce{CO} abundances in chemical nonequilibrium.
The trend of decreasing \kzz\ with lower pressure is qualitatively consistent with the forward modeling of \kzz\ results \citep[e.g.,][]{mukherjee2022}.
Our results therefore support that \kzz\ is altitude dependent in cold giant planet atmospheres.

However, the absolute values of \kzz\, as well as the quantitative difference in \kzz\ among species, should be interpreted with caution.
Both the derived quenched pressures and \kzz\ scale exponentially with temperature.
Yet, the quenched pressures fall in the region (P
$\ge$ 20 bars) in which the temperature-pressure profiles are extrapolated and are not well constrained by the data. 
To illustrate this sensitivity, we provide a diagram complementary to Figure \ref{fig:kzz} and discussion in Appendix \ref{app:kzz} to demonstrate how a small change in the quenched pressure of \ce{NH3} leads to a large variation in the inferred \kzz\.
With these caveats acknowledged, we regardless tentatively interpret our results as showing a qualitatively trend of lower \kzz\ at lower quenched pressures.
We further discuss the caveats of our models in Section \ref{sec:kzz_discuss2} and compare the derived \kzz\ to forward models and previous studies in Section \ref{sec:kzz_discuss}.

\section{Discussion}\label{sec:discussion}
\subsection{Comparison to the results in \citealt{voyer2025}} \label{sec:voyer2025}
\begin{table}[]
\begin{tabular}{lcc}
\hline
\multicolumn{1}{c}{}                & This Work                 & Voyer et al.(2025) \\
\multicolumn{1}{c}{}                &  (NIRSpec+                &                        \\
\multicolumn{1}{c}{}                & NIRCam)                & (MIRI-LRS)                       \\ \hline
Mass (\mjup)                             & 7 $\pm$ 1& $0.75^{+0.24}_{-0.16}$ \\
T$_{\mathrm{eff}}$ (K)                          & 386 $\pm$ 4             & 343 $\pm$ 11                             \\
Radius (\rjup)                               & 0.82 $\pm \,0.02$  & 1.12 $\pm$ 0.02                   \\
log(g)                              & 4.42 $\pm$ 0.01           & 3.10 $\pm$ 0.1                             \\
$\log(X_{\ce{H2O}})$                          & -3.39 $\pm$ 0.04            & -3.19 $\pm$ 0.03                           \\
$\log(X_{\ce{CH4}})$  & -3.51 $\pm$ 0.04            & -3.58 $\pm$ 0.02                             \\
$\log(X_{\ce{NH3}})$   & -4.58$^{+0.05}_{-0.04}$   & -4.74 $\pm$ 0.03                     \\
$\log(X_{\ce{CH4}}$/$X_{\ce{H2O}})$  & -0.12 $\pm \,0.06$& -0.39 $\pm$ 0.04 \\
$\log(X_{\ce{NH3}}$/$X_{\ce{H2O}})$ & -1.19 $\pm\, 0.06$& -1.55 $\pm$ 0.04
\\ 
Absolute C/O & 0.76 $\pm\, 0.03$  & 0.34 $\pm$ 0.06 \\ \hline
\end{tabular}
\caption{Comparison of the PICASO retrieval results based on NIRSpec and NIRCam data to the retrieval results of \citet{voyer2025} which is based on the MIRI-LRS data.}\label{tab:voyer2025}
\end{table}

\citet{voyer2025} (hereafter V25) published the JWST MIRI/LRS and photometric data of \target.
Because it is computationally expensive to include the MIRI LRS spectrum that triples the covered wavelength range from 1--5\um\ in the NIRSpec data out to 1--16\um\, we opted to focus \rev{instead on a comparison of the MIRI data and NIRCam photometry reported in this paper to the extrapolation of our best-fit retrieval model spectrum.}
In this section, we compare the best-fit retrieval results from the MIRI wavelength range and discuss their implications to the atmospheric properties of \target.

\begin{figure}
    \centering
    \includegraphics[width=1.0\linewidth]{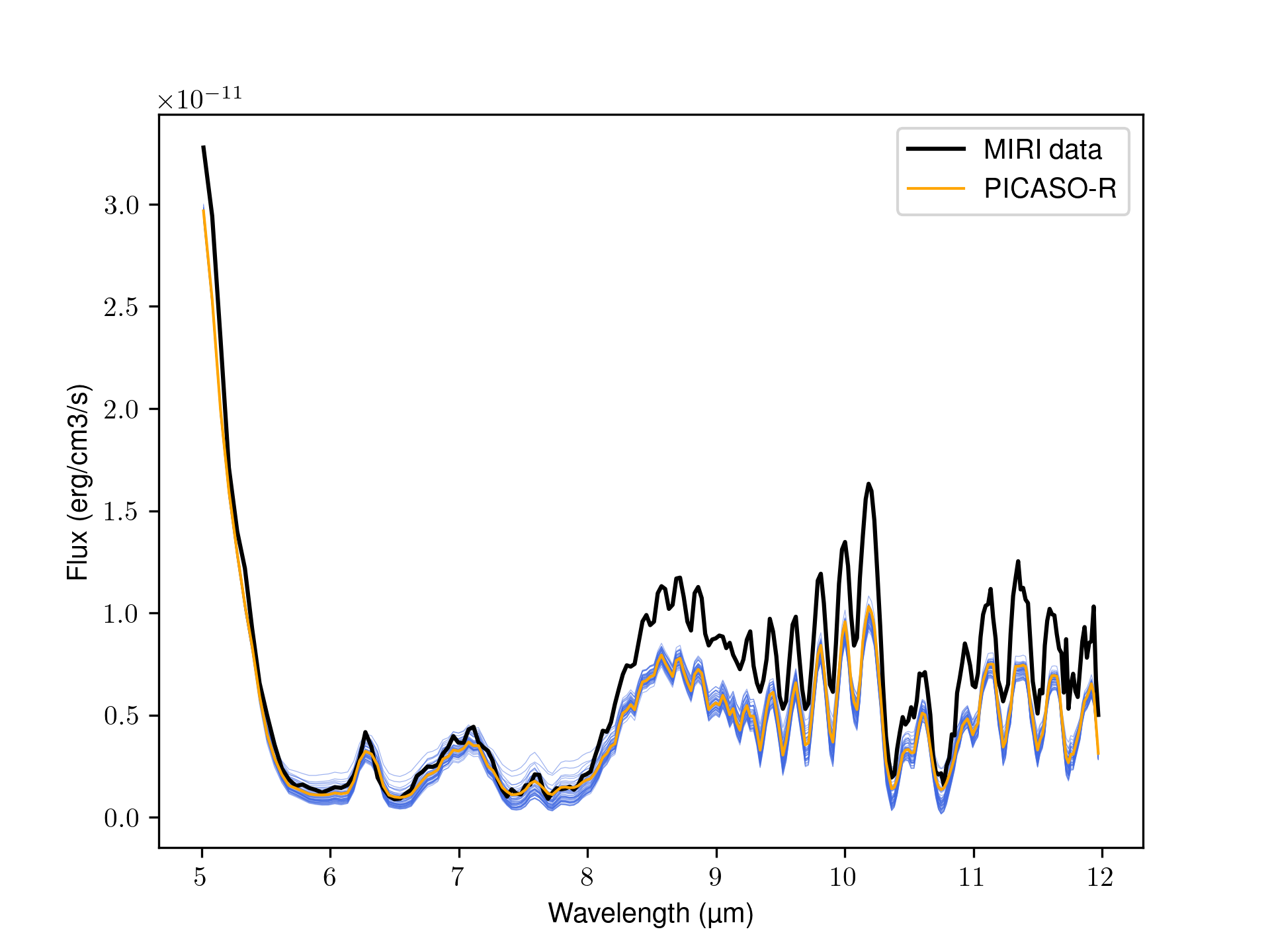}
    \caption{The MIRI/LRS spectrum (black line), which is not included in the atmospheric retrieval modeling, is about 30\% brighter than the best-fit retrieval model (orange line). The 50 randomly sampled retrieval models from the posterior distribution are plotted in semi-transparent blue lines.}
    \label{fig:miri}
\end{figure}

In Figure \ref{fig:miri},  we compare the PICASO retrieval spectrum to the MIRI data. We convolved the model spectrum with a four-pixel wide box convolution kernel and binned the model spectrum to the data resolution. We found that the retrieved model spectrum extrapolated at MIRI wavelengths shared similar spectral features with the MIRI data, but the MIRI flux is about 10\% and 40\% higher shortward of 7\um\ and longward of 8\um\ respectively.
The MIRI-LRS 5--11.7\um\ spectrum corresponds to the 0.1--1\,bar region based on the contribution function in Fig. \ref{fig:tp}.
The lower-than-observed PICASO retrieval spectrum at wavelengths longer than 6\um\ can be explained by two factors: (1) the TP profile of the PICASO retrieval is colder than that of \citet{voyer2025} in the 0.1--1\,bar range (see Fig. \ref{fig:tp}) and (2) the \ce{NH3} abundance of the PICASO retrieval is higher than that in \citet{voyer2025} (see Table \ref{tab:voyer2025}).

In Table \ref{tab:voyer2025}, we list the key retrieved parameters of PICASO retrieval and V25. The PICASO retrieval prefers a smaller radius, higher effective temperature, higher surface gravity,  and a higher mass than V25.
Assuming \target{} shares the same age as the host white dwarf, the PICASO retrieved mass matches better to the Sonora Bobcat evolutionary models than V25, while the retrieved effective temperature and radius in V25 align better to the evolutionary models.
We argue that our solution provides a more realistic mass estimate and further discuss the retrieved temperature and radius in Section \ref{sec:radius}.  

In terms of molecular abundances, the retrieved \ce{CH4} of the PICASO retrieval is similar to that of V25, while the retrieved \ce{NH3} and \ce{H2O} differ by about -0.16 dex and +0.20 dex respectively.
We note that we compared the bottom-layer  \ce{NH3} abundance of the two-layer \ce{NH3} solution of \citet{voyer2025} in Table \ref{tab:voyer2025} because the NIRSpec spectrum probes a higher pressure than the MIRI spectrum.
The retrieved \ce{CH4}, the key carbon-bearing species, are similar while the retrieved \ce{H2O}, the key oxygen-bearing species, are different between the two instrument's retrievals. 
The difference in \ce{H2O} abundances leads to different C/O ratios --- 0.76 in this study and 0.34 in V25.
It is unclear whether the difference in the retrieved \ce{H2O} reflects a systematics bias between the retrievals or indicates that the \ce{H2O} abundance, as well as the C/O ratio, changes with altitude.
We note that V25 explored retrieval with a two-layer \ce{H2O} model and did not find significant Bayesian evidence that supported a non-uniform \ce{H2O} profile.
A future joint retrieval applied to both MIRI-LRS and NIRSpec data is warranted to further investigate the potential non-uniform water abundance.

\subsection{Comparison to evolutionary models and forward models: implication of small retrieved radius}\label{sec:radius}

Given the age of 2 $\pm$ 0.5 Gyr and the measured bolometric luminosity in Section \ref{sec:lbol} , Sonora Bobcat evolutionary models \citep{marley2021} predict that the radius of \target is 1.08$ \pm 0.02$ \rjup and the corresponding temperature is $346 \pm 4$\,K.
The retrieved radius and temperature are $0.82 \pm 0.02$ \rjup\ and 386$\pm4\,$K, respectively.
The mismatch between the radius expected from forward models of the evolution and retrieved deserves attention. 
Radii as low as 0.8 \rjup\ are expected only for the most massive subsolar [M/H] brown dwarfs \citep[e.g., Figure 5 of ][]{davis2025}.
In this section, we propose a few possible reasons that could explain the smaller-than-expected retrieved radius, which is correlated  to the hotter-than-expected effective temperature.

The first is a potential systematic bias in the retrieval frameworks. Multiple retrieval studies of brown dwarfs and exoplanets found that the retrieved radius to be smaller than the evolutionary model's predicted radius \citep[e.g.,][]{zalesky2019,gonzales2020,kitzmann2020,burningham2021,hood2023,rowland2024}.
It is therefore possible that our retrieval frameworks are also subject to solutions that bias toward a small radius and a high temperature.
The second possible explanation is that the observed spectrum is a combination of two cooler atmospheric components. In this scenario, the most extreme temperature difference between the two components would be 386 K (the PICASO retrieval result) and 233 K, assuming that the cooler component exhibits zero flux in the 3--5\um\ and the sum of emission from the two components equal to the evolutionary model luminosity:
\begin{gather*}
    L_{bol,evolution} = L_{1} + L_{2} \\
    R_{evol}^2 T_{evol}^4 = R_1^2 T_1^4 + (R_{evol}^2-R_1^2) T_2^4\\
    \mathrm{From\, PICASO\text{-}R,\, R_1 = 0.82 R_J\, \&\, T_1 = 386 K}, \\
    \mathrm{then}\,   T_2 = 233 K 
\end{gather*}
However, both the coldest brown dwarf WISE 0855 and Jupiter spectra are bright in 3--5\um\, so the cooler atmospheric component would likely have to have a temperature much hotter than 233 K in a \target heterogeneous atmosphere scenario.

It remains unclear whether model biases or atmospheric heterogeneity are primarily responsible for the apparent inconsistency between the inferred radius and that predicted by evolutionary models.
Previous studies have found planetary radii that are consistent with evolutionary models \citep[e.g.,][]{kothari2024} as well as those that are inconsistent with evolutionary models \citep[e.g.][]{line2017,zalesky2019,zalesky2022,zhang2025} and the root cause remains undetermined. 
For our case, perhaps the complex irradiation history of the atmosphere as the stellar primary evolved has resulted in a disconnect between the atmospheric structure (and thus the inferred bulk properties of the object) and what would be expected for a normally evolving isolated object. 
We suggest that future studies consider such possibilities as the source of the various observed radius discrepancies is sought.  

\subsection{Subadiabatic temperature gradient}\label{sec:tpgradient}

\begin{figure}
    \centering
    \includegraphics[width=1.0\linewidth]{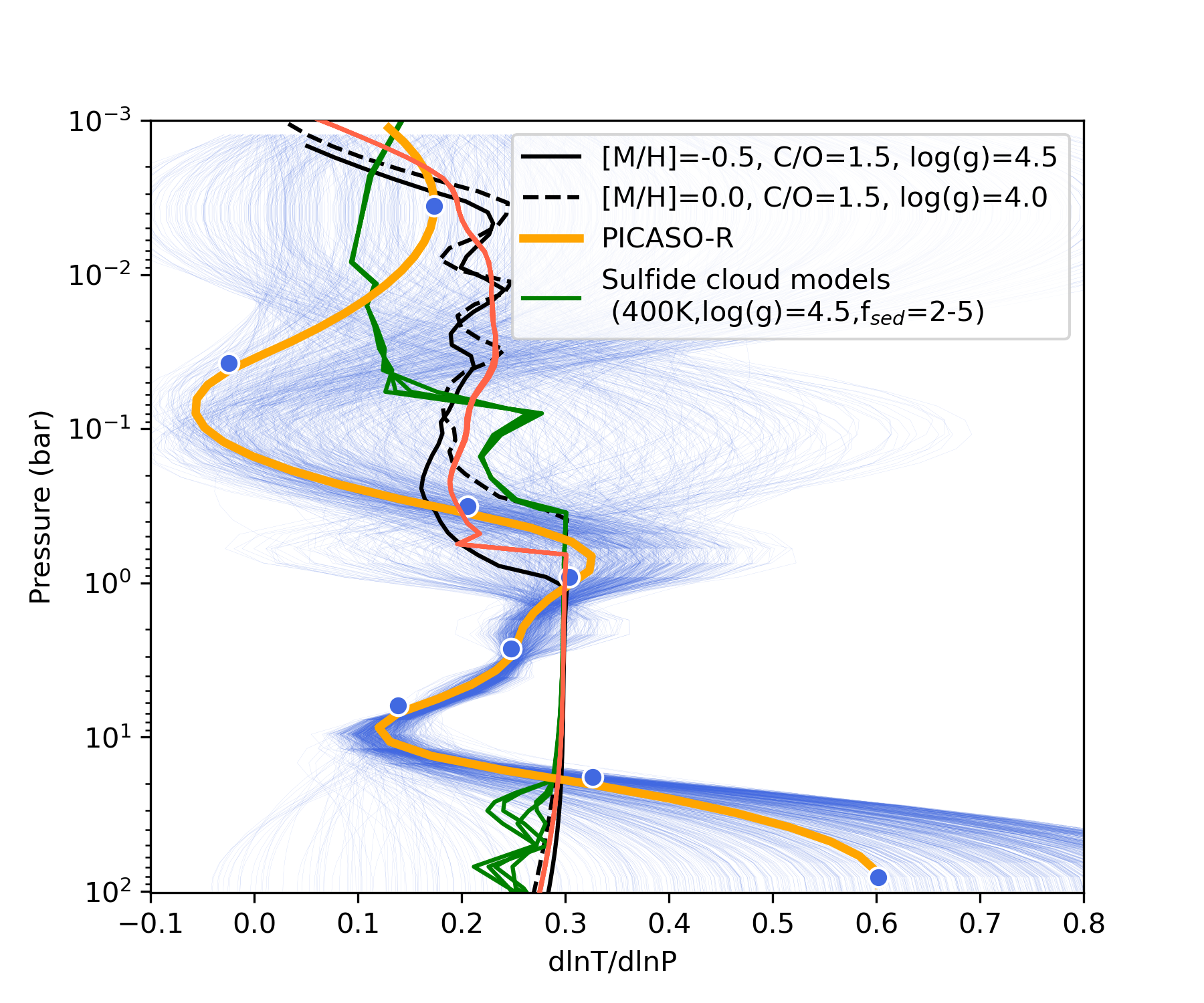}
    \caption{The temperature gradient (dlnT/dlnP) of the best-fit retrieval solution, which is plotted in an orange line, are sub-adiabatic (dlnT/dlnP $<$ 0.3) in the 1--10 bar range, which could imply a low cooling rate at depth. See Section \ref{sec:tpgradient} for more detailed discussion. At P $>$  $\sim$10\,bar, the TP profiles are not well constrained.
    The blue semi-transparent lines show the temperature gradient calculated from 1000 TP profiles randomly sampled from the retrieval posterior distribution.
    The blue dots are the TP knots in the PICASO retrieval.
    For comparison, we also plot the TP profiles of cloudless (black dashed and solid lines, Sonora Bobcat, \citealt{marley2021}) and cloudy (green solid lines, \citealt{morley2012}) forward model with an effective temperature of 400\,K similar to \target{}, but with different metallicity (${\rm[M/H]} = [-0.5, 0.0]$), gravity ($\log(g) = [4.0, 4.5]$), or sulfide cloud structures ($f_{sed} = [2,3,5]$). Both cloudless and cloudy models show that the temperature gradients closely follow an adiabat (i.e., dlnT/dlnP $\sim$ 0.3) at pressures greater than 1 bar.}
    \label{fig:tpgrad}
\end{figure}

In Figure \ref{fig:tpgrad}, we compare the temperature gradient dlnT/dlnP of the PICASO retrieval solution to that of two suites of self-consistent atmospheric models, the cloudless Sonora Bobcat models \citep{marley2021} and the sulfide cloud models \citep{morley2012}. The sulfide cloud models include Cr, KCl, ZnS, \ce{Na2S}, and MnS cloud species and include the cloud latent heat effect that reduces the temperature gradient.

We find that the temperature gradients are subadiabtic with dlnT/dlnP lower than 0.3, the dry adiabat value.
The subadiabatic temperature gradient in the 1--10 bar region does not agree with the self-consistent atmospheric models, which predict that convection should be the main energy transport mechanism at this depth and the temperature gradient should closely follow a \ce{H2}-\ce{He} dry adiabat (see the black lines in Figure \ref{fig:tpgrad}). 
We also find the subadiabat gradient is lower than that of the sulfide cloud models \citep{morley2012}(see the green lines in Figure \ref{fig:tpgrad}).
However, our finding of a subadiabat is not new - other retrieval studies of brown dwarf and giant planet atmospheres have arrived at similar conclusions \citep[e.g.,][]{line2017}. 
Previous studies claim that the subadiabat is near the edge of the pressure range probed by observational data, so the subadiabat could be sensitive to the parameterization of the TP profile.
For our case, we find that the subadiabatic profile manifests itself across the pressure region in which the TP profile is tightly constrained (see Figure \ref{fig:tp}).
While the atmospheric sub-adiabatic temperature profile changes the emergent flux and thus the cooling rate, it is unlikely to extend into the deep interior because of the high opacity and the fact that the planet has cooled from an initially hotter state. 


 \subsection{Interpretation of the sub-solar elemental abundances }\label{sec:subsolar}

\begin{figure}
    \centering
    \includegraphics[width=1.1\linewidth]{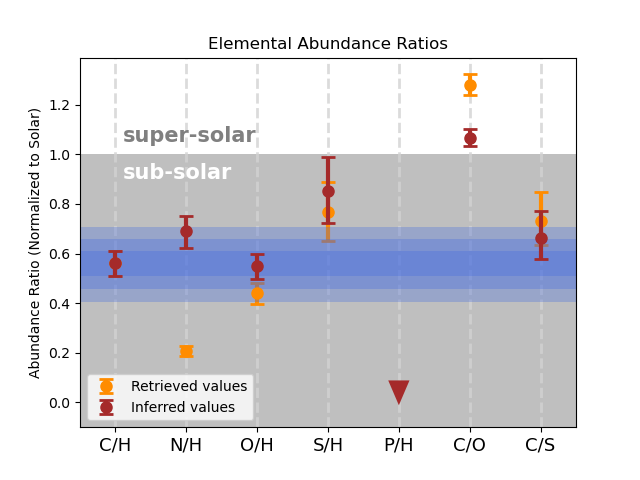}
    \caption{Our retrieved and the inferred elemental abundance ratios normalized to the solar value of \citet{asplund2021}. The P/H ratio is \rev{a 3$\sigma$ upper limit}. The blue-shaded regions denote the 1-,2-, and 3-$\sigma$ interval range of the retrieved C/H ratio. After correcting for non-detected molecules and atmospheric processes as described in Section \ref{sec:subsolar}, the inferred C/H, N/H, O/H, S/H, and C/S values agree to each other within uncertainties. The retrieved and inferred C/O ratios are both above the solar value even after accounting for oxygen sequestration during silicate cloud formation.}
    \label{fig:elementratio}
\end{figure}

In this section, we discuss if the retrieved elemental abundances are consistent with the solar elemental abundances, and what are the key chemical reactions and missing molecules for an accurate interpretation of the elemental abundances in the \target atmosphere.

For C/H, the JWST/G395M spectrum comprises the absorption features of all three main carbon-bearing species, so the measured C/H is a robust tracer of the intrinsic C/H, as well as the metallicity.
For N/H, chemical equilibrium models suggest that 70\% of N atoms form as \ce{N2}; 
Cloud formation models of silicate and sulfide clouds by \citet{visscher2006,visscher2010} suggest that the sequestration of oxygen and sulfur due to cloud formation lower the gas-phase O/H and S/H by around 20\% and 10\% respectively.

Based on the \ce{N2} abundance at the quenched pressures estimated by chemical equilibrium models,  we expect that 30\% of N is in \ce{NH3}, so the actual N/H ratio is $0.7 \pm 0.06$ and the absolute N/C ratio to be $0.3 \pm 0.04$ (c.f. $1.19 \pm 0.16$ solar), consistent with relative elemental abundance ratio of \citet{asplund2021}.
Considering the sequestration due to cloud formation and the presence of \ce{N2}, the inferred elemental abundance ratios and the retrieved abundance ratios in Section \ref{sec:elementratio} are presented in Figure \ref{fig:elementratio}.
The consistency across the inferred C/H, S/H, N/H, and O/H validate the relative abundance ratio in exoplanetary atmospheres and our chemical model correctly predicts the major C-, N-,O-, and S-bearing molecules.

\subsection{C/O and C/S ratios in the context of star and planet populations}\label{sec:hypertia}
The carbon-to-oxygen (C/O) ratio has been postulated as a tracer for a gas giant's formation location because the C/O ratio of the gas in a protoplanetary disk is expected to increase with outer disk radius as volatiles freeze out \citep[e.g.,][]{oberg2011}. 
Studies such as \citet{molliere2022} suggest that the actual scenario could be complicated because of the presence of the disk's chemical evolution and pebble drift and evaporation.
In addition to carbon- and oxygen-bearing species, sulfur volatiles also freeze out in disk, providing a complementary diagnostic \citep[e.g.,][]{turrini2021,crossfield2023} to further test and validate the use of elemental abundance ratios as tracers of a planet's formation location.
However, the detection of sulfur-bearing molecules like \ce{H2S} and \ce{SO2} were inaccessible until the launch of the JWST. With a JWST observation that covers \ce{SO2} and \ce{H2S}, we can also obtain sulfur elemental abundances. 

\begin{figure*}
    \centering
    \includegraphics[width=1.0\linewidth]{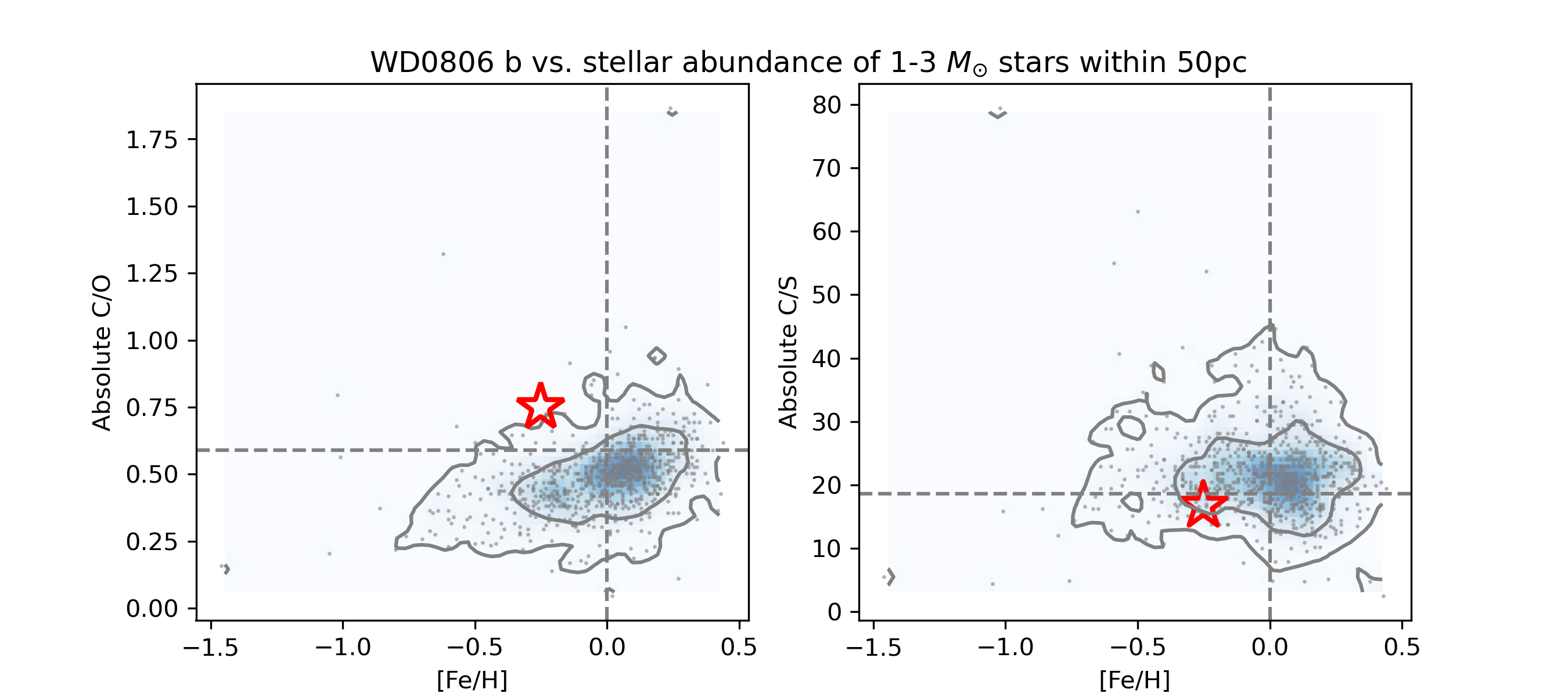}
    \caption{\textbf{Left panel}: a comparison of the absolute C/O ratio of \target{} (red star) to the nearby ($<$ 40pc) solar neighborhood population from the Hypatia catalog \citep{hinkel2014} shows that the elevated C/O ratio is consistent with nearby stellar chemical abundances within 2$\sigma$. \textbf{Right Panel:} same as the left panel, but for the C/S ratio. The C/S ratio of \target is within 1$\sigma$ to the nearby stellar chemical abundances.}
    \label{fig:hypertia}
\end{figure*}

As one of the widest orbit exoplanet companion to a white dwarf, the WD0806b atmospheric composition provides an interesting study case if a giant planet composition reflects the formation and evolution history of this rare exoplanet system.
Because the white-dwarf abundances are unknown, we resort to comparing the WD0806 to the nearby ($<$50pc distance) stellar population members that have masses similar to the progenitor mass of 1--3\,$M_{\odot}$ in the Hypertia catalog \citep{hinkel2014}.
In the left panel of Figure \ref{fig:hypertia}, the C/O ratio is about 1--2$\sigma$ higher than the stellar abundance, assuming 20\% oxygen sequestration due to silicate cloud formation.
The C/S ratio is found to be consistent with the stellar population within $1 \sigma$ level.
Based on a qualitative picture of planet formation,  the near-solar abundance ratios of C/S and C/O imply that WD0806 formed beyond the key CO, \ce{CO2}, and \ce{H2O} snowlines, which are less than or around 10-50 au for a 2-3 $M_{\odot}$ progenitor star \citep[e.g.,][]{kennedy2008,qi2015}.
Our measurement provides another important data point for future statistical analysis of exoplanet elemental abundance ratio, which has been limited size so far.

Next, we compare the reported C/O ratio of \target{} with those of other exoplanet companions. 
Our reported C/O ratio of $0.76\pm 0.03$ is higher than that of for COCONUT-2b \citep[$0.44\pm 0.012$;][]{zhang2025},  an exoplanet companion that shares a similar temperature of $483^{+44}_{-53}$K with a sub-solar or near-solar metallicity.
Comparing to the reported C/O ratios in around 30 exoplanet companions in \citet{xuan2024,hoch2023}, the measured C/O ratio of WD 0806 b is the second highest measured among planetary mass companions, with the highest reported C/O ratio of 1.39 for ROSS 458 C in \citet{gaarn2023}.
We note that our retrieved super-solar C/O and a 7 \mjup\ for \target deviates from the reported  trend among C/O and companion mass and separation reported in \citet{hoch2023}.

 \subsection{Comparison of the derived \kzz to previous results} \label{sec:kzz_discuss2}

In this section, we briefly describe the commonly used method in the community to infer the \kzz\ and compares our results to other studies. To investigate the impact of non-equilibrium chemistry on spectra across different \kzz\ values, several forward model grids \citep[e.g.,][]{karalidi2021, mukherjee2024} include \kzz\ as a free parameter. In contrast to the approach presented in this study, as described in Section \ref{sec:kzz_discuss}, these other models assume a vertically constant \kzz\. Such model grids are commonly used to infer \kzz\ values for the atmospheres of brown dwarfs and exoplanets \citep[e.g.,][]{leggett2012, kuhnle2025}.

Several studies have utilized or modified forward model grids to constrain \kzz\ in brown dwarfs with temperatures comparable to our target ($<500$ K), where the 3--5\um\ photosphere likely resides in the convective zone:
\begin{itemize}
\item \citet{miles2020} adjusted the CO abundance in the Sonora Bobcat models to fit ground-based 4.5--5.1\,\um\ spectra, estimating \logkzz\ values of 5.3--8.5 for three brown dwarfs with \teff{} below 500 K.
\item \citet{leggett2012} found best-fit \logkzz\ values of 5.5--6 for the $\sim$500 K brown dwarf UGPS J072227.51-054031.2 using the \citet{saumon2008} grid, which spans \logkzz\ from 2 to 6.
\item \citet{leggett2021} manually tuned ATMOS models \citep{phillips2020} to fit the spectral energy distributions (SEDs) of seven late-T and Y dwarfs, yielding best-fit \logkzz\ values between 6.0 and 8.7.
\item \citet{kuhnle2025} reported a best-fit \logkzz\ of $2.25 \pm 0.04$ and \teff{} of $250 \pm 6$ K for the Y dwarf WISE J0855-0714, based on fitting the 1--20,\um\ JWST spectrum with Elf-Owl V1.0 models.
\item \citet{kothari2024} constrained \kzz\ by minimizing the difference between abundances derived from atmospheric retrieval and those from a grid of non-equilibrium models. They reported a best-fit \logkzz\ of $9.0^{+0.0}_{-0.5}$,cm$^2$/s for the Y dwarf WISE J035934.06-540154.6.
\end{itemize}

In summary, these studies, which assume a constant \kzz\ in an atmosphere, report a wide range of \kzz\ values that overlap with our results. Our derived \logkzz\ range of 1.7--5.6 at a \teff{} of 386 K falls along the lower boundary of the trend tentatively identified in Figure 13 of \citet{miles2020}, which suggests an increase in \kzz\ with decreasing effective temperature.

The factors driving the observed diversity in \kzz\ among current samples remain unclear. As discussed in \citet{mukherjee2024}, the influence of non-equilibrium chemistry on the thermal profile and resulting spectra is sensitive to metallicity and gravity. 
Additional spectral analyses of brown dwarfs with well-constrained metallicities and gravities ??- especially companions with age estimates --- will be critical to better understand the variation in vertical mixing in cold brown dwarf atmospheres.

\subsection{Comparison of the derived \kzz\ to forward models and the implication of low \kzz } \label{sec:kzz_discuss}

In this section, we compare the derived \kzz\ values from Section \ref{sec:kzz} to predictions from forward models and discuss the implications. A common approach in self-consistent forward models \citep[e.g.,][]{ackerman2001, mukherjee2022} is to assume that convection is the sole mechanism for transporting energy from the interior up to the top of the convective zone. Under this assumption, \kzz\ in the convective zone is calculated following the equation in \citet{gierasch1985} (see also Equation 5 in \citealt{ackerman2001}), which yields an altitude-dependent \kzz\ profile.
Using this approach and accounting for radiative feedback from opacity variations due to non-equilibrium molecular abundances, the models of \citet{mukherjee2022} predict \kzz\ values of $10^7$--$10^8$\,cm$^2$/s in the convective zone (pressures $>10$ bar) for atmospheres with properties similar to \target{} (i.e., $T_{\rm eff} = 400$ K, solar metallicity, and $\log(g) = 4.5$).
These modeled values exceed the derived \kzz\ range of $10^{1.7}$--$10^{5.6}$,cm$^2$/s from Section \ref{sec:kzzresults} by two to five orders of magnitude.

One possible explanation for this discrepancy is a systematic underestimation of the quenched temperature and/or quenched pressure. As shown in Equations \ref{eq:kzz}, the chemical timescales decrease exponentially with temperature. 
Because the derived \kzz\ is inversely proportional to the chemical timescale, it is highly sensitive to the temperature at the quench point.
A higher quenched temperature, possibly at a higher quenched pressure than inferred from our retrievals would shorten the chemical timescale and lead to higher \kzz\ estimates.
For example, in the case of \ce{CO2}, the median temperature at the quench pressure is 780 K, with a standard deviation of 20 K. If the actual temperature were 100 K higher (a 5-$\sigma$ or 13\% deviation), the chemical timescale would decrease by a factor of $10^{3.29}$, and the corresponding \kzz\ would increase by the same factor ? more than three orders of magnitude. 
Future spectroscopic observations in the 1-2\,\um\ range, which better constrain the deep temperature-pressure (TP) profile, will be crucial for refining \kzz\ estimates for \target.

Alternatively, if the low derived \kzz\ values (ranging from $10^2$ to $10^5$,cm$^2$/s between 20 and 50 bar) are accurate, this could imply the presence of a deep, detached radiative zone in the atmosphere of \target, where convection is inefficient. 
The presence of a deep radiative zone would also be consistent with the subadiabatic temperature gradient identified in Section \ref{sec:tpgradient}, a feature shown to be necessary for reproducing Y dwarf spectra in prior modeling efforts \citep[e.g.,][]{leggett2021}, and previously suggested by retrieval studies of late-T dwarfs \citep{line2017}.
Further studies of the atmospheric circulation \citep[e.g.,][]{freytag2010,showman2019, tan2022}, detached radiative zones \citep[e.g.,][]{miles2020,mukherjee2022,mukherjee2024}, thermo-compositional adiabatic convection \citep[e.g.,][]{tremblin2019} and other possible processes are needed to determine whether a radiative zone could develop at such depth under sub-solar metallicity and high C/O conditions.
 Our results that depict a pressure-dependent \kzz\ scenario warrant further study of the impact of non-equilibrium chemistry with non-uniform \kzz\ on spectra and photometry, which are fundamental in characterizing cold giant planets and inferring their bulk properties \citep[e.g.,][]{matthews2024,galiuffi2025}.

\section{Conclusions}
\target is a benchmark cold directly imaged planet system with known age. Our JWST NIRCam and NIRSpec observations unveil the fundamental atmospheric properties of this benchmark exoplanet companion. Our key findings are as follows:
    \begin{enumerate}
    \item We present the NIRCam photometry and NIRSpec G395M spectrum of \target. We do not find any obvious companion(s) around based on the NIRCam images.
    
    \item By combining our 1--5\um\ NIRCam photometry and the NIRSpec G395M with the published 5--21\um\ MIRI LRS spectrum and photometry, we estimate the bolometric luminosity of \target{} to be $\log(L/L_{\odot}) = -6.75 \pm 0.01$. Given the estimated bolometric luminosity and an age of $2\pm0.5\,$Gyr, we infer that \target{} has a mass of $7\pm 1\,$\mjup\ based on the Sonora Bobcat evolutionary model (Section \ref{sec:lbol}).

    \item Using Elf-Owl v2.0 and ATMOS++ grid models, we find that the best-fit spectra provide reduced chi-squared values of 30 and 49 respectively. The two best-fit grid models share similar radii but have different effective temperatures (358 vs 402), surface gravity log(g) (3.3 vs. 4.5), and bulk metallicity [M/H] (-0.94 vs. 0.03) (Section \ref{sec:spectral_modeling}).

    \item Using the open-source radiative transfer Python package PICASO \citep{batalha2019}, we developed a retrieval framework to characterize the atmosphere of \target with 24 parameters. In the PICASO retrieval, we include an additive and a multiplicative uncertainty parameters, $\epsilon_{back}$ and $\epsilon_{flat}$, to account for the JWST systematics that dominate the low and high SNR regions of the spectra (Section \ref{sec:picasor}). 
    
    \item The PICASO retrieval results suggest that \target has an effective temperature of $386 \pm 4$ K, metallicity [M/H] of $-0.25 \pm 0.04$, log(g) of $4.42\pm 0.04$, and an absolute C/O ratio of $0.75 \pm 0.03$. The C/O ratio is 1.3 times of the solar C/O value of \citet{asplund2021}. We obtained bounded molecular abundances of \ce{CO2}, CO, \ce{H2O}, \ce{CH4}, \ce{H2S}, and \ce{NH3}, and placed an upper limit on the \ce{PH3} abundance (Section \ref{sec:spectral_modeling} and \ref{sec:elementratio}).

    \item We used molecular abundances to derive the elemental abundance ratios C/H, N/H, S/H, O/H, C/O and C/S. Comparing to the solar abundances of \citet{asplund2021}, we find that S/H and C/H are both subsolar and consistent within 1.7$\sigma$.  
    The similar values indicate that C/H and S/H give similar estimates of bulk metallicity. With the bounded abundances of all three major carbon-bearing and the major sulfur species, we argue that C/H and S/H are better tracers of bulk metallicity compared to other metrics (i.e., N/H, O/H, or (C+O)/H) for weakly irradiated and cold exoplanet atmospheres similar to \target{} (Section \ref{sec:elementratio}).

    \item We developed a new chemical analysis framework to derive altitude-dependent eddy diffusion coefficients for \ce{CO},\ce{CO2}, \ce{NH3} (Section \ref{sec:kzzmethod}). We found that \ce{CO2} quenched at 22 bars while \ce{NH3} and \ce{CO} are both quenched at around 47 bars. The inferred log(Kzz [cgs]) for \ce{CO2}, \ce{NH3},\ce{CO} are $1.7 \pm 0.6$, $3.2^{+1.3}_{-0.6}$, and $5.6^{+0.5}_{-0.3}$ respectively. 
    For the first time, our study provides observational evidence that vertical mixing strength decreases with higher altitude in exoplanet atmospheres.
    We caution that the inferred \kzz\ values are sensitive to the extrapolated TP profiles. 
    Thus, while the qualitative trends are robust (i.e., \kzz(\ce{CO2}) $<$ \kzz(\ce{NH3})), the quantitative differences between \kzz\ should be interpreted with caution. (Section \ref{sec:kzzresults})
    
    \item The retrieved atmospheric C/O is among the highest among ultra-cool brown dwarfs, but the bulk C/O is likely lower considering 20-30\% of the oxygen is sequestered in the silicate and possible water cloud formation. We therefore interpret that the elevated atmospheric C/O is not significantly different than the the spread of C/O abundances among solar neighborhood. We also find that C/S ratio share similar values with nearby stellar populations. (Section \ref{sec:hypertia})

\end{enumerate}
Our results demonstrate the rich atmospheric physical and chemistry processes unveiled by the high precision JWST data, including the chemical composition, CNOS(P) elemental abundance, temperature-pressure structure, vertical mixing, and bulk atmospheric properties. As one of the most peculiar planetary system with a projected orbital distance of 2500au, our detailed atmospheric characterization will be valuable for comparisons with the atmospheric properties of other similarly cold objects to further test if atmospheric composition could be useful in understanding the formation and evolution history of exoplanetary systems.

\section*{Acknowledgment}
\rev{We thank the anonymous referee for the constructive comments and suggestions that significantly improved the clarify and reproducibility of this work.}
B.L., T.R., and T.G. would like to acknowledge the funding support from NASA through the JWST NIRCam project though contract number NAS 5-03127 (M. Rieke, University of Arizona, P.I.).
M.D.F. is supported by an NSF Astronomy and Astrophysics Postdoctoral Fellowship under award AST-2303911. 
B.L. would like to thank Kevin Zahnle, Brianna Lacy, and Ehsan Gharib-Nezhad for the useful discussions.
This work benefited from the 2025 Exoplanet Summer Program in the Other Worlds Laboratory (OWL) at the University of California, Santa Cruz, a program funded by the Heising-Simons Foundation and NASA.
This work benefited from the URSSI Summer School funded through NASA Transformation to Open Science Training call, specifically award no. 80NSSC25K7771, ``Bringing Together Open Science and Research Software". 
This work employs \citet{openai_chatgpt} to refine the grammar, clarity, and overall readability of the manuscript.
Resources supporting this work were provided by the NASA High-End Computing (HEC) Program through the NASA Advanced Supercomputing (NAS) Division at Ames Research Center.
Some of the data presented in this article were obtained from the Mikulski Archive for Space Telescopes (MAST) at the Space Telescope Science Institute. The specific observations analyzed can be accessed via \dataset[doi:10.17909/dacd-v990]{https://doi.org/10.17909/dacd-v990}.

\appendix

\section{Corner plot of retrieval results}
\begin{figure*}  \includegraphics[width=1.0\textwidth]{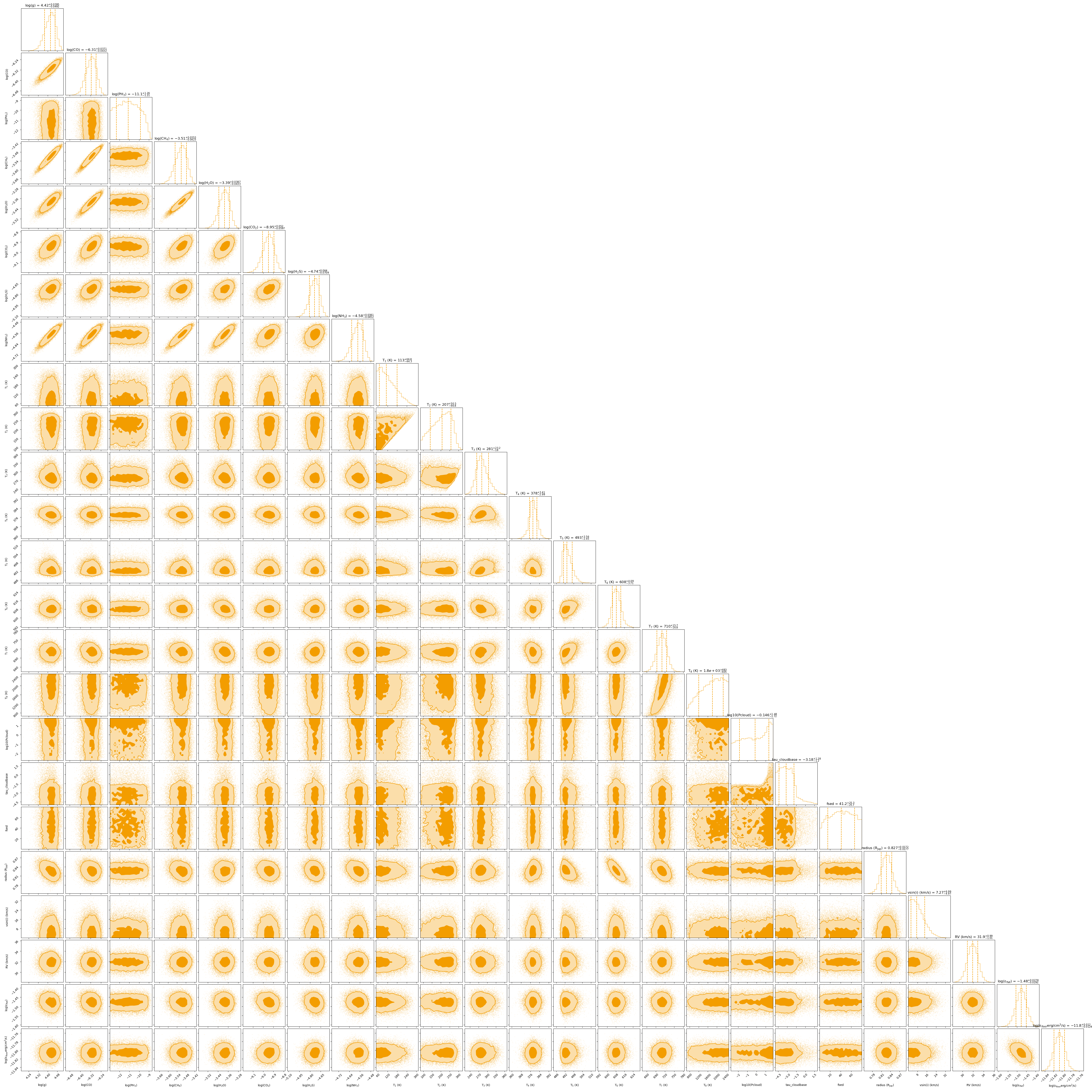}
\caption{The corner plot of the 23 parameters in the PICASO retrieval with age constraints. The marginalized distribution is shown at the top of each column with the corresponding median,16-percentile, and 84 percentiles values.} \label{fig:corner}

\end{figure*}

Figure \ref{fig:corner} shows the corner plot of the parameters used in the PICASO retrieval.

\section{Visualization of the inferred \kzz\ sensitivity on the retrieved TP profile}\label{app:kzz}
\begin{figure*}
    \centering
    \includegraphics[width=0.9\linewidth]{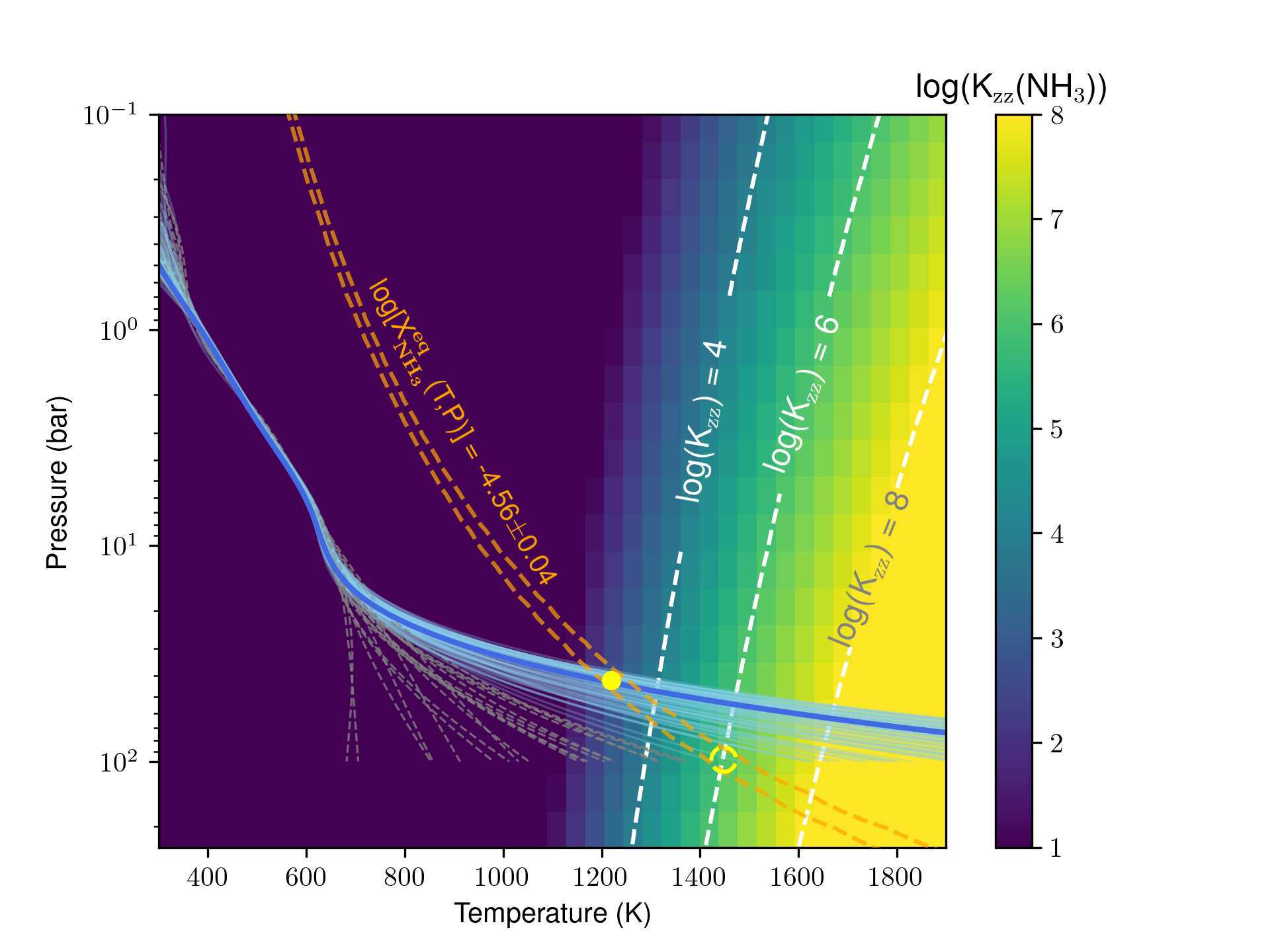}
    \caption{A colored 2-D diagram to show the corresponding derived \kzz\ as a function of temperature and pressure. The white dashed contour lines mark the \logkzz of 4, 6, and 8 respectively from left to right. The dark blue line shows the best-fit retrieved T-P profile while the light blue solid lines and dark grey dashed liens show the sampled T-P profiles from the retrieval posterior samples. The two orange dashed lines bracket the temperature-pressure region in which the retrieved \ce{NH3} abundances are equal to the chemical equilibrium abundances. The intersection between the orange \ce{NH3} contour lines and the blue T-P profile lines are the quenched pressures, which are marked with a yellow dot. The \kzz\ of the quenched pressure of the yellow dot is around $10^{3.7}$.
    If the retrieved T-P profile is cooler, the quenched pressure can move to a higher pressure and a higher temperature, such as shown with the yellow empty circle with a dashed edge, leading to a  \kzz\ that is 2 order of magnitude higher than the best-fit \kzz. The gray dashed lines are the T-P profiles that do not intersect with the \ce{NH3} contour lines and thus do not have a valid \kzz\ estimates.}
    \label{fig:nh3_tp}
\end{figure*}
The inferred \kzz\ in Section \ref{sec:kzzresults} has order-of-magnitude uncertainties, even though the retrieved abundance is precise to percent level.
In this section, we aim to explain how the derived \kzz\ relates to the retrieved TP profiles and molecular abundance using \ce{NH3} as an example.
In the complementary of plotting of the inferred \kzz\ and quenching pressures in Figure \ref{fig:kzz}, we also provide another plot to visualize the \kzz\ drastically increase with higher quenching pressure along the T-P profile.
In Figure \ref{fig:nh3_tp}, we first set up a grid of temperatures and pressures.
We assumed that mixing length is equal to the pressure scale height ($L=H$) and use the \ce{NH3} chemical timescale equation in \citet{zahnle2014} and equation \kzz\ $ = L^2/\tau_{chem}$ to calculate the \kzz\ at each grid point.
The calculated \kzz\ value is plotted in the colored T-P diagram in Figure \ref{fig:nh3_tp}.
Based on the \kzz\ diagram, we can see that \kzz (\ce{NH3}) only reaches $10^6$ or higher when \ce{NH3} is quenched at a temperature of 1400 K or hotter within a 0.1-200 bar range. 

We plotted the sampled T-P profiles in light blue and gray dashed lines (their difference will be explained later) from the posterior samples of retrieval and highlight the best-fit TP profile with the dark blue line.
At each T-P grid point, we used  the Photochem model and the median C/O  and metallicity value in Table \ref{tab:element_ratios} to calculate the chemical equilibrium \ce{NH3} abundance.
We then calculated the iso-abundance contour line at which the equilibrium \ce{NH3} abundances $\log(X^{eq}_{\ce{NH3}}$) are equal to the 1-$\sigma$ above and below the median retrieved \ce{NH3} abundances $\log(X_{\ce{NH3}}) = -4.56 \pm 0.04$.
The yellow point in Figure~\ref{fig:nh3_tp} represents the quenched pressure, which is the point where \ce{NH3} is in chemical equilibrium with an abundance value that is the same as the retrieved value. 

As shown in the Figure, the retrieved T-P profiles is well constrained by the NIRSpec spectrum in the 1--10\,bar region and has large uncertainties at P$>$10 bars. 
Depending on where the T-P profile intersects with the \ce{NH3} abundance contours, the yellow point location on the T-P diagram varies and leads to an order of magnitude difference in the derived \kzz.  
For example, the dashed empty yellow circle in the plot represents a scenario where the cold T-P profile that intersects the \ce{NH3} abundance contour at a higher pressure and leads to higher derived \kzz\ compared to that of the solid yellow point.

We also point out that some TP profiles sampled from the posterior distribution of retrieval results, which are plotted in grey dashed lines, do not intersect with the \ce{NH3} abundance contours within the pressure grid of our models. In this case, quenching pressure estimates are not available and no \kzz\ can be derived, so these T-P samples are excluded from the calculation of \kzz.

\clearpage
\bibliography{main}{}
\bibliographystyle{aasjournal}



\end{document}